# Spin- and time-resolved photoelectron spectroscopy and diffraction studies using time-of-flight momentum microscopes

G. Schönhense and H.-J. Elmers

*Institut für Physik, Johannes Gutenberg-Universität Mainz, D-55099 Mainz, Germany*

**ABSTRACT**

Momentum microscopy (MM) is a novel way of performing angular-resolved photoelectron spectroscopy (ARPES). Combined with time-of-flight (ToF) energy recording, its high degree of parallelization is advantageous for photon-hungry experiments like ARPES at X-ray energies, spin-resolved and time-resolved ARPES. This article introduces into the technique of ToF-MM and illustrates its performance by selected examples obtained in different spectral ranges. In a multidimensional view of the photoemission process, spectral density function $\rho(\boldsymbol{k},E_B)$, spin polarization $\boldsymbol{P}(\boldsymbol{k},E_B)$ and related quantities of circular dichroism in the angular distribution (CDAD) are part of the 'complete experiment', a concept adopted from atomic photoemission. We show examples of spin-resolved valence-band mapping in the UV, VUV, soft- and hard-X-ray range. Spin mapping of the Heusler compounds $Co_2MnGa$ and $Co_2Fe_{0.4}Mn_{0.6}Si$ at hν=6eV prove that the second compound is a half-metallic ferromagnet. Analysis of the Tamm state on Re(0001) using VUV-excitation reveals a Rashba-type spin texture. Bulk band structure including Fermi surface, Fermi-velocity distribution $\boldsymbol{v_F}(\boldsymbol{k},E_F)$, full CDAD texture and spin signature of W(110) have been derived via tomographic mapping with soft X-rays. Hard X-rays enable accessing large $k_\parallel$-regions so that the final-state sphere crosses many Brillouin zones in $k$-space with different $k_z$. At hν=5.3keV this fast 4D mapping mode (at fixed hν) revealed the temperature dependence of the Fermi surface of the Kondo system $YbRh_2Si_2$. Probing the true bulk spin polarization of $Fe_3O_4$ at hν=5keV proved its half-metallic nature. The emerging method of ToF-MM with fs X-ray pulses from free-electron lasers enables simultaneous valence, core-level and photoelectron diffraction measurements in the ultrafast regime.

# I. INTRODUCTION

Momentum microscopy (MM) is a new and powerful way of performing angle-resolved photoelectron spectroscopy (ARPES) and X-ray photoelectron diffraction (XPD). ARPES in general and its soft- and hard-X-ray variants (SXARPES and HARPES) are the methods of choice for electronic structure analysis. Their importance and impact are steadily growing, being fueled by the discovery of topological states and exciting electronic properties of quantum materials[1,2,3], for which high spatial resolution (NanoARPES)[4] becomes increasingly crucial. Accessing the spectral function of *in operando* devices can reveal changes in a materials' electronic structure by field-effect gating or current passage[5]. *In-situ* and *operando* methods are important for understanding interface creation and interface evolution in different environments[6]. Related to this new class of ARPES experiments is the electronic response on mechanical strain as visible in the stunning changes of the electronic structure in twisted graphene[7].

Observation of XPD dates back to the pioneering work of Siegbahn et al.[8] and Fadley and coworkers[9,10]. Later, several groups developed XPD as an avenue toward fast and effective recording of structural information[11,12,13,14]. Strongholds of XPD are its capabilities to analyze adsorbate geometries[15], its close relation to photoelectron holography[16,17,18,19] and even a spin-polarized version has been developed by the Fadley group[20,21]. Its element specificity and, hence, high site specificity can be exploited as unique fingerprint of atomic sites in compounds[22]. By combining ARPES and XPD in a single experiment, electronic and structural information can be obtained quasi-simultaneously and at identical conditions (kinetic energy of the photoelectrons, size and position of the probing spot and probing depth)[23].

The last decades have seen dramatic advances in the technical performance of photoemission experiments, both on the source and detector side. Storage rings for synchrotron radiation reach physical limits of emittance and brightness. Beamlines with high resolving power toward $10^5$ are available at many storage rings, especially in the soft- and hard-X-ray range. Free-electron lasers (FELs) and laboratory-based higher-harmonic-generation (HHG) sources provide fs pulses for time-resolved photoemission experiments in the ultrafast regime.

On the detector side, hemispherical analyzers (HSAs) are the working horses of photoelectron spectroscopy. Today, HSAs have reached excellent performance with high energy resolution (down to < 1 meV)[24,25,26,27] and high angular resolution (equivalent to a momentum resolution of 0.003 Å$^{-1}$)[28]. In the classical mode of operation [($E_{kin}$,θ)-mode] the exit plane of a HSA hosts a 2D detector, which can record an energy interval of typically 10% of the pass energy and an angular interval of +/-7° in the high-angular-resolution mode. Larger angular ranges can be recorded using wide-angle lenses or sequentially via sample rotation.

The MM family comprises several types of instruments, the energy-dispersive type using either a tandem arrangement of two hemispherical analyzers[29,30,31] or a single hemispherical analyzer[32,33]. The key feature of all MMs is their large solid-angle acceptance, exploiting the full-field-imaging properties of a cathode lens. Combination of the MM concept with time-of-flight detection has led to the ToF-MM[34]. More recently, a hybrid instrument of single hemisphere plus ToF section has been developed[32,35]. The functioning scheme of a MM exploits a basic principle of optics, namely the fact that the backfocal plane of an objective lens hosts a reciprocal image (in mathematical language, also called Fourier image). In charged-particle optics this reciprocal image is nothing else than the lateral momentum distribution, briefly termed k-image. MMs record huge intervals in ($E_{kin}$,***k***) parameter space without sample rotation or angular scanning.

ToF-MM is the youngest member in the family of photoemission instruments. This development is based on the time-resolving imaging technique originally developed for the ToF-PEEM[36,37]. Naturally, time-of-flight energy recording needs pulsed photon sources. Thanks to their perfectly-periodic time structure, synchrotron radiation, pulsed lasers, higher-harmonic-generation (HHG) sources and fs-pulsed sources like free-electron lasers (FELs)[38,39] or the upcoming attosecond facilities[40,41] fit to the ($E_{kin}$,$k_x$,$k_y$) recording scheme of a ToF-MM. A natural expansion of the technique is a time-resolved variant of ToF-MM, utilizing fs-pulsed infrared/optical pump and X-ray probe pulses to track the time-evolution of the state as it evolves following optical excitation. The method was recently established at the FEL FLASH at DESY, Hamburg[42,43,44]. Further ToF-MM projects are just being launched at the European XFEL[45] in Schenefeld and LCLS-II[46] in Stanford.

In this article we present the status and future potential of *time- and spin-resolved ToF-MM* and the related phenomena of *circular (and linear) dichroism in the angular distribution* (CDAD, LDAD). The experiments are discussed in terms of a multidimensional description of the photoemission process and the concept of the 'complete experiment' (adopted from atomic and molecular photoemission), comprising intensity, dichroism and spin measurements. After introducing the experimental technique, examples are presented for various spectral ranges. The 4$^{th}$ harmonic of a Ti:sapphire laser oscillator (h$\nu$= 6.05 eV, 100 fs pulses) was used for spin-resolved band mapping of two ferromagnetic Heusler compounds. The high pulse rate of the laser (80 MHz) yielded count rates of almost 1 MHz in the spin detector. A spin-resolved band mapping in the full 2$\pi$ solid angle ($\theta$=0-90°, azimuth $\phi$=0-360°) and 1.5 eV energy interval below $E_F$ could be completed in about ½ hour after in-situ film growth. Spin measurements using synchrotron radiation in the vacuum ultraviolet (VUV) at BESSY II are exemplified by the Tamm surface state of Re(0001), showing Rashba signature (ground-state spin polarization). Optical spin orientation $P_{Fano}$ induced by circularly-polarized soft X-rays and the spin component perpendicular to the reaction plane $P_{Mott}$, also arising from spin-orbit interaction, have been studied at beamline P04 (h$\nu$= 300 – 1700 eV) at PETRA-III. The first spin-resolved ToF-MM measurements in the hard X-ray range (h$\nu$= 5 keV) at beamline P22 probed the bulk spin polarization of magnetite, addressing the question whether $Fe_3O_4$ is indeed a half-metallic ferromagnet. Tomographic-like *k*-space mapping using soft X-rays with variable polarization is shown to capture Fermi-surface, Fermi-velocity, band dispersions, dichroism and spin. For the Kondo system $YbRh_2Si_2$ *k*-space tomography at h$\nu$= 5.3 keV in a novel, 4D recording scheme revealed a change of the Fermi surface and band dispersions when varying the temperature. As an outlook we briefly address the emerging method of fs time-resolved photoemission using FEL excitation, showing an example recorded with the high-energy X-ray ToF (HEXTOF) at FLASH (DESY, Hamburg).

## II. EXPERIMENTAL CONCEPT

### II.A. Multi-dimensional description of the photoemission process: complete experiment, orbital mapping and wavefunction reconstruction

Before discussing the ToF-MM technique, it is worthwhile to consider the 'multi-dimensionality' aspect of the photoemission process and its implications for the recording technique. The power of photoemission is that it 'probes the electronic structure of a sample/material'. The central quantity is the *spin-dependent spectral density function* $\rho_{\uparrow\downarrow}$ ($E_B$,***k***), defined in energy-momentum space. For a static system the parameter space is 4-dimensional with the exception of strictly 2D systems like surface states, adsorbates or some layered materials. The spin information is essential for magnetic materials (examples in III.A and III.E). In materials showing phase transitions, the (lattice) temperature $T_l$ is an important parameter influencing the spectral density. As example we will show the $T_l$-dependence of the Fermi surface in the Kondo system $YbRh_2Si_2$ (III.D).

The rapidly-growing field of time-resolved experiments addresses another important parameter, the intrinsic time scale $\tau$ of a dynamic process or a transient electronic state (example in III.F). In pump-probe experiments we encounter a complex scenario, depending on pump fluence, light polarization and pump photon energy. In the perturbative limit, the pump pulses generate single-electron excitations, decaying with characteristic lifetimes in the sub-ps range. In the high-fluence regime, the large amount of energy deposited by an intense fs pump pulse leads to highly non-equilibrium states with increased temperatures of the electron and spin system along with transient changes of the band structure. Here, we adopt the simplified description of non-equilibrium, assuming that individual subsystems can be described by (transient) temperatures, e.g. electronic and spin subsystems being described by $T_e(t)$ and $T_s(t)$, respectively. Such multi-temperature models, while not without flaws[47,48,49], are still commonly used to interpret time-resolved data. Tracking the time-evolution of the electronic structure allows one to disentangle different relaxation pathways like energy transfer from the hot electrons to the optical and acoustic phonon system and the spin system, all happening on different time scales. By such an approach interaction mechanisms and coupling strengths between different degrees of freedom can be determined. Hence, for the description of a pump-pulse-driven phase transition in a time-resolved photoemission experiment we have to consider the response of the spectral density in a multidimensional parameter space $\rho_{\uparrow\downarrow}(E_B,\boldsymbol{k},T_e,T_l,T_s,\tau_i)$, where $\tau_i$ denote the intrinsic time scales of several relaxation channels. Using ToF-MMs such experiments have become feasible at FELs[42-44] and HHG sources[50]. The parameters $T_e$, $T_s$, $T_l$ (when discussing the time evolution of phonons, one often distinguishes between temperatures of weakly coupled acoustic modes and strongly coupled optical phonon branches) and $\tau_i$ are interlinked in a complex way, since charge, spin and lattice are not in thermal equilibrium and thermalization between different subsystems proceeds on different time scales[51, 52, 53, 54]. Moreover, all of the parameters $T_e$, $T_l$, $T_s$, $\tau_i$ in general depend on the absorbed energy density, as do the electronic bands, energy isosurfaces etc.. This multi-dimensional aspect, going beyond multi-temperature models is an <u>inherent property of the electronic system</u>, independent of the spectroscopic technique.

When claiming that photoemission probes the electronic structure, we have to keep in mind that the photoelectron signal does not reflect the spectral function $\rho$ itself but results from the interaction operator and the Green function in the one-step description of angle-resolved photoemission[55]. According to Fermi's Golden Rule, the observed signal at a certain ($E_B,\boldsymbol{k}$) is proportional to the squared transition matrix element of the photon operator between the initial and final-state wavefunctions[56,57]. Symmetry selection rules in this matrix element can lead to disappearance of signals from bands with certain symmetry for a given photon polarization. Such effects are accessible via variation of the photon polarization[58]. These symmetry selection rules lay the cornerstone of a family of experiments probing *circular or linear dichroism,* CDAD[59] or LDAD[60]. They are no ground-state properties, the dichroism asymmetries $A_{CDAD,LDAD}$ ($E_B,\boldsymbol{k}$) are defined in the same parameter space as the spectral density (example in III.C). New discoveries like the *time-reversal dichroism* (TRDAD[61]) and a new appearance of an *intrinsic linear dichroism* (iLDAD[62]) in layered materials indicate a large future potential of dichroism measurements.

A serious aspect of the matrix-element effects concerns the extracted spin information. Only in special cases like ferromagnets with negligible spin-orbit interaction can photoemission measure the true ground-state spin polarization. Such a favorable example is spin-resolved hard-X-ray photoemission from magnetite, discussed in III.E. In the general case the photoemission process itself alters or even generates spin polarization. A prominent example of the latter is the optical spin orientation (Fano effect[63]) by circularly polarized light, first verified for a solid (GaAs) in a pioneering experiment by Pierce and Meier[64]. Later it has been shown that even unpolarized or linearly-polarized light can yield highly-spin-polarized electrons from unpolarized atoms[65,66] and nonmagnetic solids[67]. This spin-polarization component arises due to spin-orbit interaction and resembles the polarization in Mott scattering[68]. In III.C we show examples for the *Fano* and *Mott spin components* in photoemission from W(110). A previously-overlooked relation between these spin components and the CDAD is discussed in Ref. 69.

The (vector) quantity **P**($E_B$,***k***) observable in photoemission in general carries a superposition of several signatures: A possible ground-state contribution (in magnetic and Rashba systems) is altered by 'intrinsic' processes caused by the transition matrix element[70,71,72] and by 'extrinsic' processes like the spin-dependent transmission and rotation of the spin vector during transport through a magnetic material[73] or the matching of the spinor functions at the solid-vacuum interface[74] or a possible depolarization through spin-flip scattering at defects or surface contaminations.

Let us look into the quantitative details of a *'complete' experiment* in the sense as introduced by Kessler[75] and Cherepkov[76] for gas-phase photoemission. Recording all experimentally-accessible quantities: photoelectron intensity, angular distribution and three components of the spin vector can yield the full quantum-mechanical information on the photoemission dynamics of an atom. We can transfer this model to solid-state photoemission, keeping in mind that the numbers of partial waves in the initial and final states can be very large. Unlike for a free atom, we do not have the means to derive a full set of matrix elements and phases purely from the experiment. Exceptions are cases like the π-band of graphite, where the pure $Y_1^0$ symmetry leads to a restricted set of final-state partial waves[77]; for further discussion of this aspect, see Ref. 78. On the other hand, the full band structure, various forms of dichroism (CDAD, LDAD, TRDAD) and three spin components, each defined in 5- or 6-dimensional parameter space contain a plethora of information. Here, the question arises whether such a 'complete' experiment will be feasible at all or remains a theoretical concept. At the end of this article, a tentative answer will be given on the basis of a quantitative estimate of required signal strength, detector performance and acquisition time. We will consider the acquisition time needed for an experiment recording the full band structure and texture of dichroism and spin with their time dependences in an ultrafast process like a pump-pulse driven phase transition.

The concept of the 'complete' photoemission experiment, introduced 4 decades ago[75], aims at the determination of a full set of quantum-mechanical parameters, which govern the photoemission process of free atoms of molecules. These are the transition matrix elements and phases of outgoing partial waves. Since then, our understanding of gas-phase photoionization has dramatically improved. By recording photoions in coincidence, the CDAD can be measured in a fourfold-differential manner[79]. These developments, mostly using the COLTRIMS technology[80], are beyond the scope of the present review. Recently, several groups pursued a related approach in condensed matter photoemission: Momentum-, energy- and time-resolved ToF-MM data is used to reconstruct the molecular orbitals [81,82,83,84] or the Bloch wavefunction itself [85,86,87,88,89 90]. Full-field k-recording is ideally suited for this goal, because the real-space image I($r_x$,$r_y$) is recovered via a 2D Fourier transform of the *k*-resolved photoemission intensity I($k_x$,$k_y$). Since this is one of the most exciting future aspects of ToF-MM, we will briefly address these first results in III.F. Although exemplified for excitonic wave functions in organic films and few prototypical layered semiconductors, the approach is very general and will provide insights into the Bloch wavefunctions of many relevant crystalline solids.

### *II.B. The time-of-flight momentum microscope with imaging spin filter*

MMs belong to the family of cathode-lens type electron microscopes, like the well-known photoemission electron microscope (PEEM). The lens system of an MM is optimized for best resolution in *k*-space, unlike a PEEM, which aims at optimum real-space imaging. Tusche et al.[30] showed a resolution of 0.004 Å$^{-1}$, hence MMs can compete with the best values of angular / *k* resolution of ARPES hemispherical analyzers (HSAs)[28]. The key advantage of MMs is that there is no restriction of the range of emission angles θ, up to kinetic energies of 70 eV. In a MM θ translates into a simultaneously observable $k_\parallel$-range, according to $k_{\parallel,max}$=0.512 sinθ$_{max}$√$E_{kin}$ (with $k_{\parallel,max}$ in Å$^{-1}$, $E_{kin}$ in eV). Since the momentum pattern exhibits a linear and largely achromatic $k_\parallel$-scale, a MM can be considered as a magnifying glass in *k*-space. Further electron lenses in the microscope project an energy-filtered momentum image with variable magnification to the detector plane. The energy filter is a key element; it can be either energy dispersive (hemispherical analyzers) or time-of-flight dispersive.

A ToF-MM detects the photo-emitted electron intensity $I(E_B, k_x, k_y)$ as a function of parallel momentum $k_x$ and $k_y$ and binding energy $E_B$[34]. Figure 1 illustrates the functioning principle. The first imaging element (objective lens) of the zoom optics behind the sample is a so-called cathode lens, of which the sample itself is an integral element. This lens forms an achromatic momentum image in its backfocal plane (BFP), see inset (b). This image is magnified to the desired size by a two-stage zoom optic comprising several lens groups (not shown). The photoelectron distribution is dispersed in time-of-flight by a field-free low energy ToF drift section and recorded in an $(k_x, k_y, t)$-resolving delay-line detector[91] (DLD 1). A movable and size-selectable field aperture in the first Gaussian image plane allows a precise definition of the analyzed region-of-interest (ROI) down to the sub-micrometer range (see Fig. 6 in Ref. 35). By switching lenses to the real-space imaging mode (PEEM mode), this aperture and its position on the sample surface can be viewed on the detector. The sample is mounted on a He-cooled hexapod manipulator for precise six-coordinates adjustment $(x, y, z, \theta_x, \theta_y, \phi)$.

The efficiency of the three-dimensional recording architecture of ToF-MMs (and angular-resolving ToF spectrometers[92,93]) is defined by the time-resolving image detector. Delay-line detectors[94,95] or the upcoming solid-state pixel-array detectors[96,97] are characterized by time resolutions in the 50-200 ps range. Hence, for ToF-based experiments not only the pulse length but also the pulse period of the photon source is crucial because the ratio between duration and period determines, how many time slices can be resolved in the time gap between adjacent photon pulses. In Sec. III we show examples using a UV-laser with 12.5 ns period (80 MHz pulse rate), synchrotron radiation with 800 ns period (BESSY II in single-bunch mode) and with 192 ns period (PETRA III in 40-bunch mode), as well as FEL radiation with 1 µs pulse period, but only 5000 pulses per second (FLASH). The resulting numbers of resolvable time slices will be stated and briefly discussed for each of the cases.

*Imaging spin filters* act like spin-selective electron mirrors and can analyze many points of a 2D electron distribution simultaneously. The spin selectivity results from spin-orbit (or exchange) interaction in low-energy electron scattering. The spin asymmetry is quantified by the Sherman function $S = (I\uparrow - I\downarrow)/(I\uparrow + I\downarrow)$ with $I\uparrow$ and $I\downarrow$ denoting the scattered intensities for an incoming beam, completely polarized parallel and antiparallel to the spin-quantization axis. The efficiency of a single-channel spin detector is given by its figure-of-merit $FoM_{single} = S^2 I/I_0$ (with $I/I_0$ denoting the reflectivity). Multichannel spin detection was pioneered by Kolbe et al., who captured the $(E_{kin}, \theta)$ spin pattern behind the exit of a conventional hemispherical analyzer,[98] and Tusche et al., who analyzed ferromagnetic domain patterns in real-space images[99] and later the spin texture in $k$-images[30].

The *effective figure-of-merit* increases with the number $N$ of resolved points in the spin-filtered image $FoM_{eff} = N \times FoM_{single}$. With an imaging electron optics, $N = 10^3 - 10^4$ 'points' on the spin-filter surface can be analyzed simultaneously. A third 'coordinate' is added by parallel energy recording via capturing the time-of-flight $\tau$ using a delay-line detector. The imaging spin filter in a ToF-MM thus yields 3D data arrays $P(E_{kin}, k_x, k_y)$, as sketched in Fig. 1(c). The $(k_x, k_y)$-region represents a sectional plane through the BZ. The width of the usable $E_{kin}$ interval can comprise up to several eV, depending on the type of spin filter. Suga and Tusche discuss many types of spin detectors and compare values for $FoM_{eff}$ (see Fig. 28 in Ref. 100). The 2D filters reach almost $FoM_{eff} = 100$ and the 3D filters can yield 1-2 orders of magnitude higher $FoM_{eff}$ values, all referenced to an 'ideal' single-channel detector with $FoM = 1$.

The spin filter integrated into a ToF-MM consists of the second ToF branch (90°-branch, DLD 2) and the retractable spin-filter crystal. Several types of spin-filter crystals have been characterized and used[101, 102, 103, 104, 105], more are proposed[106, 107]. A spin- and angle-resolved (spin-ARPES) spectrometer with a Co film as spin filter coupled with a laboratory-based UV laser has been developed by the Berkeley group[103]. Many experiments use Ir(100) with a pseudomorphic monolayer of Au, because of its almost infinite lifetime in UHV[102]. The measurements shown in Fig. 1(e) and III.A were done with the 'classical' surface W(100), which is characterized by a broad spin-asymmetry profile of > 2 eV, a very high reflectivity up to several % and the complete absence of any visible mosaicity or poly-crystallinity (as observed for other spin-filter materials). This spin filter is operated at 45° scattering angle and 27 eV scattering energy; the plane of incidence is (010). Its corresponding $FoM_{single}$= 5×10$^{-3}$ (definition, see Ref. 108). Inserting the resolution values of 0.03 Å$^{-1}$ and 20 meV from Ref. 109, one can estimate the $FoM_{eff}$ for the example shown in Sec. III.A. The momentum disc at $E_F$ comprises $N≈$1700 resolved ($k_x,k_y$)-pixels. Thanks to parallel energy recording in the ToF mode, this yields $N>10^5$ resolved ($E_B,k_x,k_y$)-voxels. We thus arrive at $FoM_{eff}≈$ 400, referenced to the energy and $k$-space interval of Fig. 2 (being relatively small because of the small photon energy, hν=6eV). The spin quantization axis of the measurement (component **P** perpendicular to the drawing plane of Figure 1(a)) points along the easy magnetization axis **M** of the sample. Reversal of **M** was done in-situ by approaching strong permanent magnets to the thin-film samples.

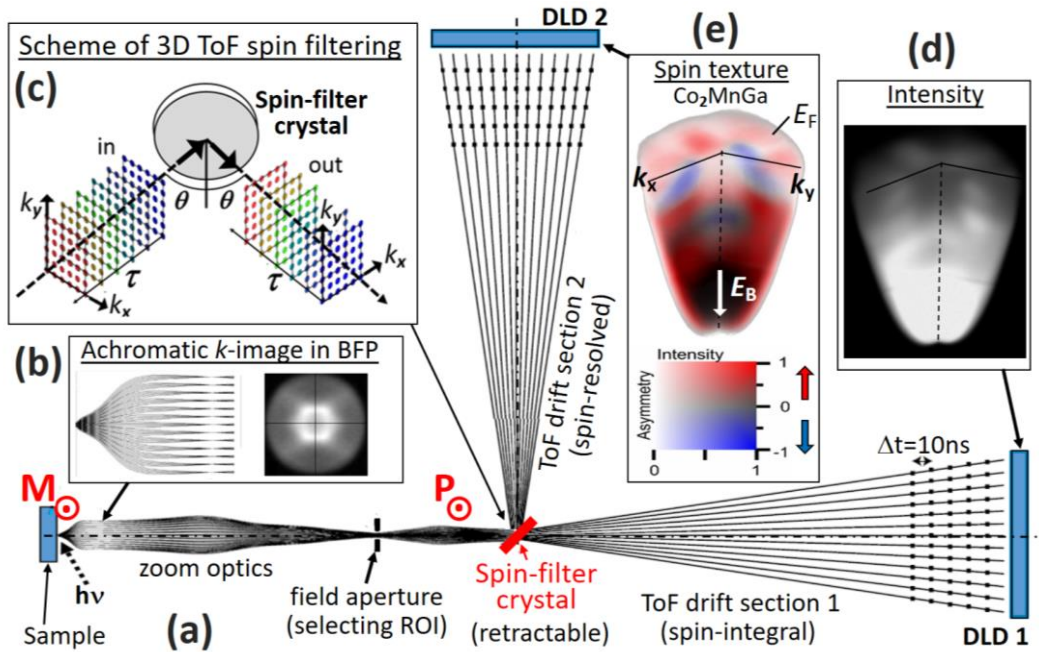

**FIG. 1.** Illustration of the functional principle of a spin-filtered time-of-flight momentum microscope. (a) Schematic overview of the electron-optical setup, comprising a lens system with objective lens and two-stage zoom optics with movable aperture in the first Gaussian (real space) image plane, selecting the region of interest (ROI). The insets describe (b) the formation of a momentum image in the backfocal plane (BFP) and (c) the scheme of ToF spin filtering. The spin quantization direction is perpendicular to the scattering plane (drawing plane), as indicated by the vector **P**. For the study of ferromagnetic spin textures, the magnetization vector **M** of the sample is parallel to **P** and can be reversed by a magnetic field. The (x,y,t)-resolving delay-line detectors in the spin-integral (DLD 1) and spin-resolving branches (DLD 2) capture the 3D data arrays of intensity and spin in a large solid-angle interval above the sample, without any scanning or sample rotation. Typical experimental data arrays of intensity $I$ ($E_B,k_x,k_y$) and spin polarization **P**($E_B,k_x,k_y$) are shown in (d) and (e), respectively. The 3D color code in (e) denotes majority (red), minority (blue) as well as unpolarized (grey) partial intensities. The data array forms a paraboloid, confined by the photoemission horizon, corresponding to an emission angle of 90° and by the cutoff at the Fermi energy. Diameter and depth of the paraboloid are defined by the kinetic-energy range and imaged $k$-interval. The examples (d,e) correspond to excitation of the Heusler compound Co$_2$MnGa with hν = 6.05 eV photons; the depth of the paraboloid is 1.9 eV and its diameter at $E_F$ is 1.4 Å$^{-1}$.

The spin-texture patterns were determined for each ($E_B$,$k_x$,$k_y$)-voxel from two measurements with reversed spin polarization or asymmetry. There are several ways of facilitating this reversal, which are exemplified in the following sections. For *ferromagnetic* samples the magnetization is reversed; examples in Figs. 1(e), 2 and 6. For excitation with *circularly polarized photons* the optical spin orientation (Fano component) is reversed by switching the photon helicity; example Fig. 4(l). If the initial spin polarization cannot be reversed (e.g. in Rashba systems), the sign of the asymmetry function needs to be switched; example in Fig. 3. This can be done by *changing the scattering energy* between two working points with extremal positive and negative asymmetry. Both reflectivity and spin asymmetry depend on scattering angle and energy. Hence, the algorithm retrieving the P($E_B$,$k_x$,$k_y$)-arrays must consider this dependence. For further details on the evaluation procedure for spin-texture arrays in imaging spin filters, see Ref. 110. A second way of reversing the scattering asymmetry utilizes *exchange scattering at ferromagnetic films*[103,106]. Here the magnetization of the ferromagnetic film in the spin filter is reversed. In favorable cases, the symmetry properties of the spin polarization texture can be exploited for deriving the spin asymmetry[78].

The local pressure in the spin filter region determines the lifetime of the surface; the W crystal used in Fig. 2 required a cleaning flash once per hour. The quality of the spin filter surface is monitored by the signal-to-background ratio of the observed momentum image. Immediately before data recording the spin-filter crystal is flashed. The fast 3D-recording ensured that a complete spin-texture map is finished without visible degradation of the spin-filter surface.

In the measurements shown in Fig. 1(e) and III.A the high pulse rate (80 MHz) of the Ti:sapphire laser yields count rates in the spin-filter branch up to 8x10$^5$ counts per second (cps), depending on the photon polarization. Thus, a typical spin-resolved measurement (like the 3D data array in the inset Fig. 1(e)) requires only ~30 minutes of acquisition time. Owing to the high repetition rate of the laser, the average photoemission yield stayed well below one photoelectron per pulse, avoiding the serious space-charge problem of low-repetition-rate systems[111, 112]. The detection efficiency in the straight branch (spin-integrated data, DLD 1) is only limited by the maximum count rate of ~5 Mcps of the DLD.

At this point it is worthwhile considering the general situation at synchrotron-radiation sources. The vast majority of experiments requires high photon flux and do not need time structure. Hence, most synchrotron run time provides maximum bunch filling with 100 or 500 MHz, with limited time dedicated to 'few-bunch' modes for experiments that exploit the pulsed nature of the light. Corresponding periods of 2-10 ns are too short for high-resolution ToF electron spectroscopy with a sufficient number of resolved time slices. There are two loopholes out of this dilemma: Either the period of the pulses can be increased by electron-optical or even mechanical pulse picking, meaning that only a sub-period of electron bunches is selected by a fast beam blanker. Or the energy interval is reduced by a bandpass pre-filter, inserting a dispersive element into the microscope. This approach retains the full period of the photon source. Both strategies have been proven successful as discussed in Ref. 35.

The repetition rate of FELs is sufficiently low to allow for ToF analysis. The 1 MHz operation mode at LCLS-II[46] will provide excellent conditions for ToF k-microscopy. Conventional HHG sources also have pulse rates up to the 1 MHz range. The cavity-enhanced HHG sources[113,114] have pulse rates in the 60-80 MHz range, defined by the optical path length. For such sources, a bandpass pre-filter or high-pass filter is required. It is worth mentioning that HHG sources can provide circularly polarized VUV radiation, either in generation[115] or using a reflective 'waveplates'[116].

## III. SELECTED EXAMPLES

### III.A. UV-Laser ARPES with spin-resolution: Spin-texture mapping of Heusler compounds

As example for the application of spin-resolved ToF-MM in the UV spectral range, Fig. 2 shows recent results for two ferromagnetic Heusler compounds [$Co_2MnGa(100)$ and $Co_2Mn_{0.6}Fe_{0.4}Si(100)$, briefly CMG and CMFS]. This pair is particularly interesting because it shows similar electronic band structures but different band filling. A comparative study[109] together with the parent compound $Co_2MnSi$, CMS, was focused on the question how band positions and dispersion depend on the different band filling, i.e., different number of valence electrons $N_V$ in these three isostructural compounds. For this experiment, the 4$^{th}$ harmonic of a Ti-sapphire laser provided fs UV pulses with a photon energy of 6.05 eV. Binding energies of up to $E_B$= 1.9 eV are accessible with a one-photon photoemission process and the visible field-of-view in $k$-space at $E_F$ is $k_{\|max}$= 0.7 Å$^{-1}$.

According to the universal curve of the inelastic mean-free path (IMFP)[117], such low kinetic energies are characterized by a large information depth, thus giving access to bulk bands. Yet, for many cases the strongly-confined photoemission horizon at low excitation energies poses a fundamental limitation on the suitability of low-energy photons for valence-band mapping. The large unit cell of Heusler compounds [Fig. 2(k)] corresponds to a small Brillouin zone (BZ) and thus represents a favorable case for low photon energies. On the polar angular scale, the present momentum-range corresponds to the full $2\pi$ solid angle and represents 63% of the surface Brillouin zone (BZ) along the main axes $k_x$ and $k_y$. The photo-transition at hν= 6.05 eV leads to the centre of the second bulk BZ. Hence the experiment probes the Γ-X and Γ-K high-symmetry directions, defined in Fig. 2(l).

The epitaxial Heusler films were grown on MgO(100) substrates at room temperature using rf-magnetron sputtering in an Ar atmosphere at a pressure of 0.1 mbar and subsequent annealing in UHV at a temperature of 550 °C. X-ray diffraction and energy dispersive X-ray spectroscopy (both ex-situ) were employed for analysis of the film quality confirming the L2$_1$ crystal structure; for details, see Ref. 118. After finishing deposition, the samples have been transferred within few minutes under UHV conditions into the sample stage of the microscope (base pressure 2×10$^{-10}$ mbar). Then the spin mapping is immediately started so that the acquisition is finished within typically half an hour after film deposition. The CMFS film degraded visibly after few hours, whereas for the CMG-films we found a remarkably long 'lifetime', with the spin-polarized bands still being visible after several weeks.

The incident photon beam is perpendicular to the y-axis and at 22° to the x-axis of the chosen coordinate system. The samples were oriented with [011] and [0-11] crystallographic directions along the x and y-axis. The photon polarization is varied using quarter- and half-wave plates, resulting in s- and p-polarization or left (LCP) and right (RCP) circular polarization. Exploiting the polarization dependence of the photoemission intensity patterns, the symmetry groups of the observed bands can be probed [119]. These symmetry selection rules allow the separation of overlapping bands with different spin signature.

Figure 2(a) shows the intensity pattern for CMG at the Fermi energy. The ($k_x$,$k_y$)-cut clearly shows several band features crossing $E_F$. The spin texture (b) uncovers that most bands have majority character (red) but two pronounced blue features along the [011]-direction (=$k_x$), about 0.25 Å$^{-1}$ from the Γ-point, stem from a minority band. The $E_B$-vs-$k_x$ sections of intensity (c) and spin (d) reveal the dispersion of this band towards larger $k_{\|}$ with increasing E$_B$. This minority band (actually a group of three bands A-C) crossing $E_F$ clearly rules out that CMG could be a half-metallic ferromagnet. In the perpendicular azimuth [0-11] section (e) we recognize a flat majority feature (bands D, E) crossing $E_F$ at almost the same $k_{\|}$-value. This example demonstrates how the matrix-element effect can be exploited for the identification of the double-group symmetries of bands in a ferromagnet. The electric field vector of the photon beam was set to s-polarization (pointing along $k_y$), thus lifting the degeneracy of the [011] and [0-11]-directions being equivalent in the bulk crystal.

For CFMS we encounter a different situation: Here the ($k_x,k_y$)-cut at $E_F$ exhibits a structure-less intensity pattern with uniformly red spin texture. Only at $E_B$= 0.5 eV the first band features with minority character (blue) occur, Figs. 2(f,g). Note the similarity of spin patterns (b) and (g) differing in energy by 0.5 eV. The $E_B$-vs-$k_∥$ cuts (h-j) show an strongly-dispersing minority band with maximum well below $E_F$. The uniformly-red spin character at and below $E_F$ is the signature of a half-metallic ferromagnet with a minority spin gap of 0.35eV with respect to $E_F$. Fig. 2(j) shows a faint residue of a surface resonance as discussed [120] for the parent compound CMS. Lower-lying bands are hardly discernible in the intensity pattern (h), but show up more clearly in the spin patterns (i,j). The dashed horizontal lines in (j) mark an exchange-split band pair, allowing a precise experimental determination of the exchange splitting in CFMS ($\Delta E_{ex}$ = 0.72 ± 0.07 eV); for CMG and CMS the values are 0.48 ± 0.07 eV and 0.55 ± 0.10 eV, respectively.

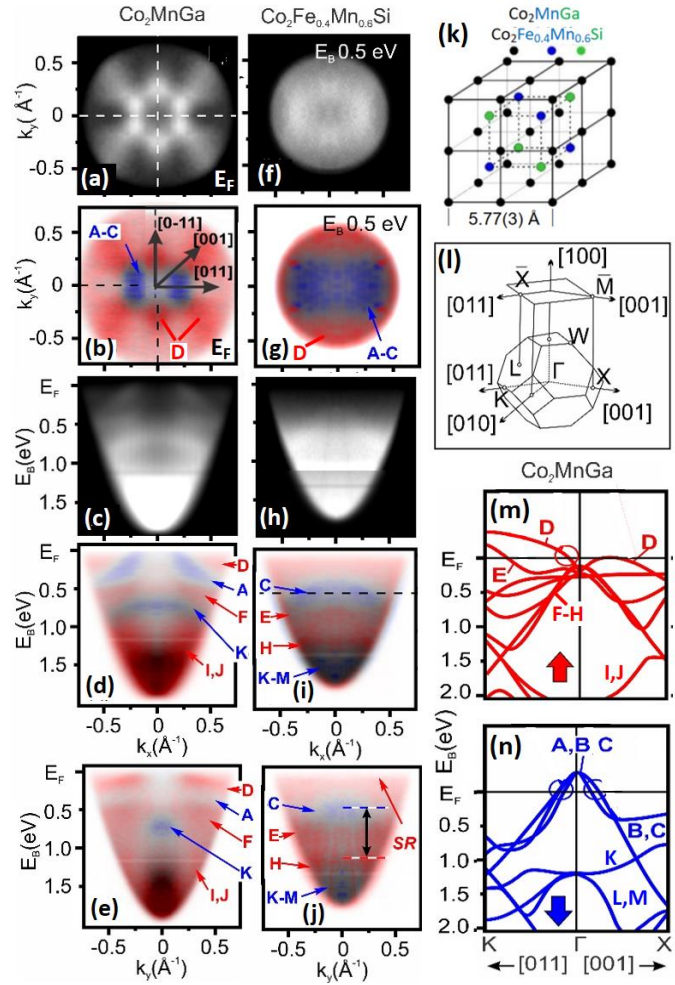

**FIG. 2.** Momentum distribution and spin-texture for in-situ-deposited Co$_2$MnGa (a-e) and Co$_2$Mn$_{0.6}$Fe$_{0.4}$Si (f-j), recorded with a ToF-MM using UV excitation (hν= 6.05 eV, s-polarized) in comparison with theory. (a,b) I($k_x,k_y$) and P($k_x,k_y$) for CMG at the Fermi energy; (f,g) same for CMFS at $E_B$ = 0.5 eV. (c-e) Band dispersions and spin texture shown as sections $E_B$-vs-$k_x$ (c,d) and $E_B$-vs-$k_y$ (e); (h-j) same for CMFS. The sections show the spin-asymmetry distribution of the topmost minority bands A, B, C and majority bands D and E (color scale as in Fig. 1(e), band labels like in theory (m,n)). (k) Large unit cell of CMG and CMFS and (l) corresponding bulk and surface Brillouin zones with high-symmetry points and directions. The high-symmetry points are at distances Γ - X = $\bar{\Gamma}$ - $\bar{M}$ = 1.1 Å$^{-1}$ and $\bar{\Gamma} - \bar{X}$ = 0.8 Å$^{-1}$. (m) Majority and (n) minority band structure for CMG calculated in DFT-LSDA code; energy scales have been adjusted with experiment. Data (a-j) from Ref. 109 and (m,n) from Ref. 120 [Reprinted with permission; copyright 2021 American Physical Society].

A comparison of the measured results (and those for CMS) with various calculations (LSDA+U, LSDA+GW, DMFT) is given in Ref. 109. Figs. 2(m,n) show the majority and minority band structure of CMG calculated by DFT-LSDA from the comprehensive comparative study of many Heusler compounds by Ma et al. [121]. Close inspection reveals that all observed band features and their polarization signature are clearly related to the calculated valence bands and band groups (labelled A-M) of bulk $Co_2MnGa$. The absolute energy positions of individual band groups have been adjusted to the experimental results at the Γ-point. The minority valence-band maximum for CMG lies only 0.15 eV above $E_F$, i.e. this compound is much closer to being a half-metallic ferromagnet than predicted by ab-initio theory.

Adjusting calculated bands to the experimental positions requires different shifts for majority and minority bands (in some cases even shifts in opposite directions). This finding is at variance with the anticipated behavior based on the rigid-band description, where just the band filling determines the position of $E_F$. The dispersion of majority and minority electrons allows one to link exchange splitting and band filling to the number of valence electrons $N_V$. We find that CMS and CFMS exhibit minority band gaps of 0.5 and 0.35 eV, opposite to the expectation from the rigid-band model. All these deviations are signatures of many-body electron correlations, which influence the size of the minority gap and the band positions. Our results indicate that current theoretical models qualitatively explain the experimental observations but show substantial quantitative differences. In view of this it is remarkable that the magnetic moments are predicted in almost perfect agreement with the experimental values. The half-metallic Heusler compounds are promising candidates for spintronics and sensors and their minority spin gap is relevant for macroscopic properties, among all for the transport properties.

### III.B. VUV ARPES: Complete analysis of Re(0001) photoemission

The broken inversion symmetry at surfaces causes a strong electric field perpendicular to the surface. This electric field provokes a lifting of the spin-degeneracy known as the Rashba-effect [28,122, 123, 124, 125]. The spin direction will then be locked perpendicularly to the momentum. Model systems, for which the Rashba-effect has been directly shown by photoemission experiments include the Shockley surface state on Au(111) [123,124,126, 127] and the surface resonance on W(110) [128, 129, 130, 131, 132, 133, 134]. As second example for the efficient spin-resolved detection scheme of time-of-flight momentum microscopy, we discuss here the Rashba-split Tamm surface state on Re(0001) [135] in more detail. The Rashba splitting of this surface state has been predicted by density functional theory [136] and quantum interference patterns revealed its pronounced surface localization [137].

The spin-resolved photoemission experiment has been performed at beamline U125/2 (BESSY) in the single bunch mode (repetition rate 1.25 MHz)[138]. In this mode the total available VUV intensity is a factor of twenty smaller than in normal multibunch mode. On the other hand, the ToF detection efficiency is two orders of magnitude higher compared to a two-dimensional detection mode.

Spin-integrated photoemission spectroscopy at different photon energies in a range from hv = 7–35 eV verifies the surface or bulk character of electronic states. As depicted in Fig. 3, the surface state, marked as SS, shows a constant parallel momentum independent on photon energy. A second fingerprint is the attenuation of its intensity with increasing photon energy, which contrasts with the surface resonances (SR). At a photon energy of 13.5 eV one observes a bulk state (B) at the Fermi level that strongly disperses with the photon energy.

To obtain the spin-resolved information, an Au/Ir(001) spin-filter crystal is inserted in the electron optical column (Fig. 1 and Ref. 110). The spin dependent reflectivity of low-energy electrons at this surface, caused by spin-orbit interaction, allows to evaluate the spin-polarization from two datasets measured at scattering energies of 10.75 eV and 12.5 eV, in the following denoted by the subscripts *l* and *h* (for low and high), respectively. The spin quantization axis perpendicular to the scattering plane is adjusted along the Γ-K direction, denoted as the *y*-direction in Fig. 3. The Au/Ir(100) spin filter crystal offers a Sherman function of +\- 0.7 at these two scattering energies [110]. The spin polarization at each image point is determined by $P_0 = (1 - \rho\psi)/(\rho\psi S_l - S_h)$, where $\psi = I_h/I_l$ is the ratio of measured intensities at each image point and $\rho = R_l/R_h$ denotes the ratio of the energy-dependent spin-integrated reflectivity $R_h$ and $R_l$ at both scattering energies. The energy dependence of $S_i$ can be assumed to be constant in the small energy interval evaluated here.

The constant-energy maps of the spin texture [Fig. 3(f-k)] reveal an outer SS' and inner SS Rashba ring manifesting the lifting of the degeneracy of spin-up and spin down band. The blue/red color code indicates the spin component along the *y*-direction. As the spin direction is perpendicular to the parallel momentum, the measured spin polarization component $P_y$ vanishes along the $k_y$-axis and obeys the relation $P_y(-k_x) = -P_y(k_x)$. The maximum observed spin polarization of the Rashba split surface state amounts to *P* = 0.7.

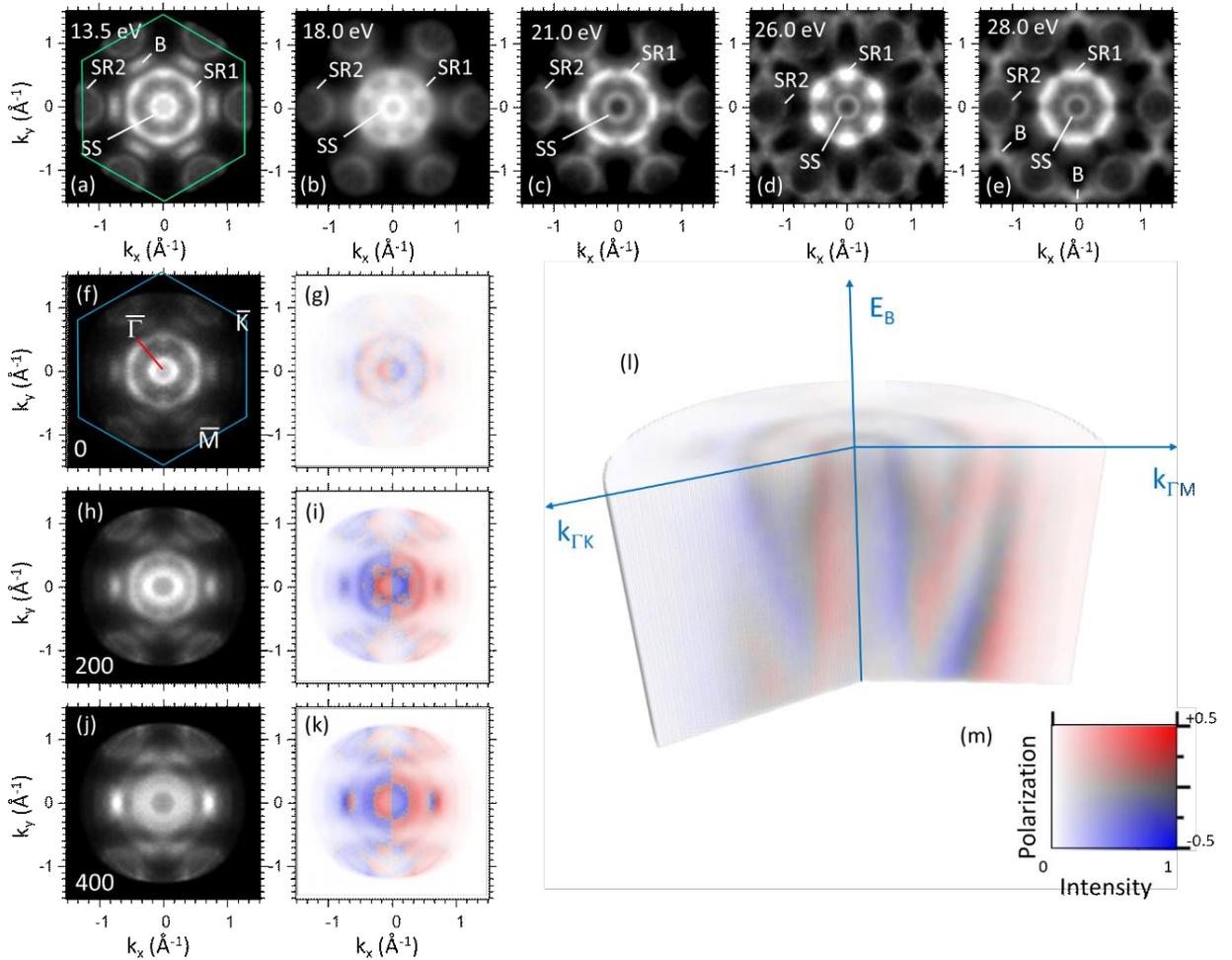

**FIG. 3.** (a-e) Constant-energy maps I($E_B$ = 0,$k_x$,$k_y$) for excitation with different photon energies as indicated. The photon energy hv = 18.5 eV corresponds to $k_z$ = 2$G_{0001}$ for normal emission, i.e., at Γ, in the free electron model. Features varying with photon energy indicate surface resonances (SR) or bulk states (B). The intensity of the surface state (SS) monotonously decreases with increasing photon energy indicating the pronounced localized character of the features emitted close to the surface normal. (f) Constant energy map I($E_B$ = 0,$k_x$,$k_y$) measured for hv = 13.5 eV after reflection at the spin-filter crystal. (g) Constant-energy spin polarization maps $P_{surf}$($E_B$,$k_x$,$k_y$). Color code for intensity and spin polarization indicated in (m). (h-k) Corresponding data for $E_B$ = 200 meV and 400 meV. (l) Cut 3-dimensional representation of the $P_{surf}$($E_B$,$k_x$,$k_y$) data array. [Adapted with permission from Ref. 135 under a Creative Commons License]

The spin-resolved band dispersion depicted in Fig. 3(l) reveals the dispersion of the surface state SS near the Γ point. The group velocity near the Fermi level amounts to –(4.8 ± 0.7) eVÅ independent on the parallel momentum direction. The Rashba spin texture causes the opposite sign for negative and positive parallel momentum. For this photon energy, the spin-resolved Rashba-split surface state can be observed up to a binding energy of 0.7 eV.

The M-point surface resonance (SR2) has a weak positive spin polarization for $k_x > 0$, corresponding to a clockwise spin direction around the M point. The bulk character of the state indicated by B has been deduced from the pronounced $k_z$-dispersion. The spin polarization of bulk state B vanishes at the Fermi level as expected but also develops a Rashba-like spin splitting at larger binding energy.

To conclude, VUV radiation leads to an extreme surface sensitivity of photo-emitted electrons. The surface sensitivity imposes a challenge for spin- and angular-resolved photoemission spectroscopy due to the notoriously low efficiency of spin detectors and necessarily long acquisition times. Spin-resolved ToF spectroscopy circumvents this challenge because of its significantly increased detection efficiency based on 3-dimensional $P(E_B,k_x,k_y)$ spin recording. The example of the Tamm surface state of Re(0001) shown in Fig. 3 demonstrates this capability, revealing the Rashba-split spin texture of this state. Additional photoemission measurements in the soft X-ray regime[135] confirmed that this surface state resides within a projected bulk band gap. The state with smaller parallel momentum (inner Rashba state) is fully separated from bulk states, whereas the Rashba branch with larger momentum hybridizes with bulk states, leading to a partial suppression of spin-momentum locking.

### *III.C. Soft X-ray ARPES: Mapping Fermi surface and velocity, circular dichroism and spin*

In the classical ARPES range (VUV) the inelastic mean free path (IMFP) of the photoelectrons is < 0.5 nm, hence VUV-ARPES is ideally suited for the study of surface states and adsorbates. With increase of photoelectron kinetic energy towards the soft-X-ray range the IMFP increases from 0.5 nm at 200 eV to 2.5 nm at 2000 eV [139]. This implies that soft-X-ray ARPES gives access to deeper-lying layers of the solid. In turn, it is possible to probe the bulk electronic structure, which is of interest e.g. in the context of transport properties. The most important aspect of SX-ARPES is its capability of enabling a tomographic-like mapping of the electronic band structure. Here we discuss how full-field imaging ToF-MM is used for a rapid mapping of electronic bands in the 4D ($E_B$,**k**) parameter space including the circular dichroism in the angular distribution (CDAD).

Soft-X-ray ARPES poses technical obstacles concerning both excitation and detection. The required energy resolution (10 meV range) and momentum resolution (~ 1% of the size of the BZ) challenge the performance of the beamline (resolving power $E/\Delta E \approx 10^5$) and the angular resolution of the electron detector (< 0.1°). Advances of the brilliance of synchrotron sources show up in a number of soft-X-ray beamlines with high resolving powers (several $10^4$), e.g., P04 [140] (PETRA III), ADDRESS [141] (Swiss Light Source), TEMPO [142] (SOLEIL), BL17SU, BL23SU, BL25SU [143,144,145] (SPring8), I09 [146] (DIAMOND) and VERITAS [147] (MAX IV). The photoemission cross sections decrease with increasing energy in the soft (and hard) X-ray regime [148], demanding high recording efficiency. The 3D recording scheme of the ToF-MM effectively counteracts the decreasing signal intensity with increasing photon energy.

The ToF-MM directly images the ($k_x,k_y$) momentum distribution in a large k-range of up to 7 Å$^{-1}$ diameter (up to 20 Å$^{-1}$ for the high-energy optics[149]) and an energy range of several eV. Capturing energy and the two transversal momentum components simultaneously in the relevant parameter range (full BZ, entire *d*-band complex), the question arises how to get access to the perpendicular momentum component $k_z$. The answer becomes obvious when considering the photo-transition in the periodic zone scheme of the bulk band structure [($E_B$,**k**)-space]. As an example, the center panel of Fig. 4 illustrates direct transitions in tungsten in the range between hν=306 and 1293 eV. The quantitative scheme reveals that this photon-energy range corresponds to final states in the 3$^{rd}$ to 7$^{th}$ repeated BZ along $k_z$.

*Energy conservation* is accounted for analogously to the Ewald sphere in a diffraction experiment. For a given photon energy hν and binding energy $E_B$ the final states of the photo-transition are located on a spherical shell with radius (for units Å$^{-1}$ and eV):

$$k_f \sim 0.512 \sqrt{h\nu - E_B + V_0^*} \times \left(\frac{m_{eff}}{m}\right) \qquad (1).$$

$V_0^*$ is the inner potential referenced to the Fermi energy. The workfunction does not enter because the scheme describes a transition in the bulk. $m_{eff}$ and $m$ are the effective mass of the electron in the final state inside the solid and the free-electron mass, respectively. The role of $k_z$ becomes immediately clear from this scheme: it is the perpendicular component $k_f^\perp$ of the final momentum vector $\mathbf{k}_f$. The electronic states in a periodic solid exhibit a strict periodicity in *k*-space. Hence the band structure is identical in all the repeated BZs and – except for different matrix elements – it is irrelevant whether the 4$^{th}$ or 20$^{th}$ repeated BZ is mapped. Only the radial integrals change when varying the period of the rapidly oscillating final-state wave function.

*Momentum conservation* is less straightforward due to two effects, the contribution of the photon momentum $\mathbf{k}_{h\nu}$ and the fact that the final state in a photo-transition is a "time-reversed LEED state". The vector $k_f^\parallel$ is conserved, when the electrons leave the sample. The photon (energy and momentum) enters into the inter-band transition and $\mathbf{k}_{h\nu}$ is completely transferred to the electron wave vector of the final state. The transfer of photon momentum to the photoelectron happens in the transition itself and leads to a rigid (vectorial) displacement of the final-state in *k*-space (see Ref. 150 and references therein). In a momentum microscope, this effect is directly observable as a shift of the momentum pattern (example Fig. 3 in Ref. 34). Given the photon impact angle of 22°, the displacement of the final-state sphere along $k_z$ and $k_y$ is

$$\Delta k_z = (2\pi\nu/c)\sin 22° \qquad \Delta k_y = (2\pi\nu/c)\cos 22° \qquad (2).$$

For the highest photon energy used at beamline P04 (hν= 1293 eV), $\mathbf{k}_{h\nu}$ shows up in substantial shifts of $\Delta k_y$ = 0.635 Å$^{-1}$ and $\Delta k_z$ = 0.237 Å$^{-1}$.

The second effect is described by Pendry's model of the 'multiple-scattering' photoemission final state[151, 152, 153, 154]. When we translate Bragg's law to k-space, we find constructive interference e.g. in normal-emission, whenever an integer-multiple of the photoelectron wavelength $\lambda_e$ coincides with the spacing of the atom planes $d_z$. With the reciprocal-lattice vector perpendicular to the surface $|G_z|=2\pi/d_z$ and the final-state electron momentum $|k_f|=2\pi/\lambda_e$, we obtain the constructive-interference condition for forward scattering

$$k_f = nG_z \qquad (3).$$

The generalization in 3D k-space is the Laue equation, describing an Umklapp-process involving an arbitrary reciprocal lattice vector **G**

$$\mathbf{k}_f = \mathbf{k}'_f - \mathbf{G} \qquad (4).$$

Pendry's multiple-scattering final state can be accounted for in a graphical scheme like Fig. 4(a) by including Umklapp processes on the final-state energy isosphere in k-space. This extension explains the observed XPD modulations and their extremely rapid variation with energy and momentum; for details, see[155].

Figs. 4(b-g) show $I(k_x,k_y)$ sections at $E_F$ and a binding energy $E_B$>0 for three selected photon energies. It is convenient to quantify the perpendicular component $k_z = k_f^\perp$ in multiples of the reciprocal lattice vector along the <110> direction, $G_{110}$=2.812 Å$^{-1}$. The momentum patterns at different photon energies are entirely different and show a strong dependence on binding energy, both being fingerprints of the 4D nature of ($E_B$,**k**)-space tomography. The $k_z$-resolution is obviously sufficient to separate band features that appear only when the spherical final-state shell intersects with the 'objects' in the BZ (i.e. constant-energy sheets, one of them being the Fermi-surface with electron or hole pockets). From the shape of the band pattern in Fig. 4(b) this section can be clearly identified as running through the centre (Γ-point) of a repeated BZ. Other patterns allow identifying further high-symmetry planes like the NHP mirror plane marked in Fig. 4(a) and mapped in Fig. 4(d). Calibrated at high-symmetry points, this soft-X-ray based tomographic-like approach provides an absolute '$k$-space ruler'.

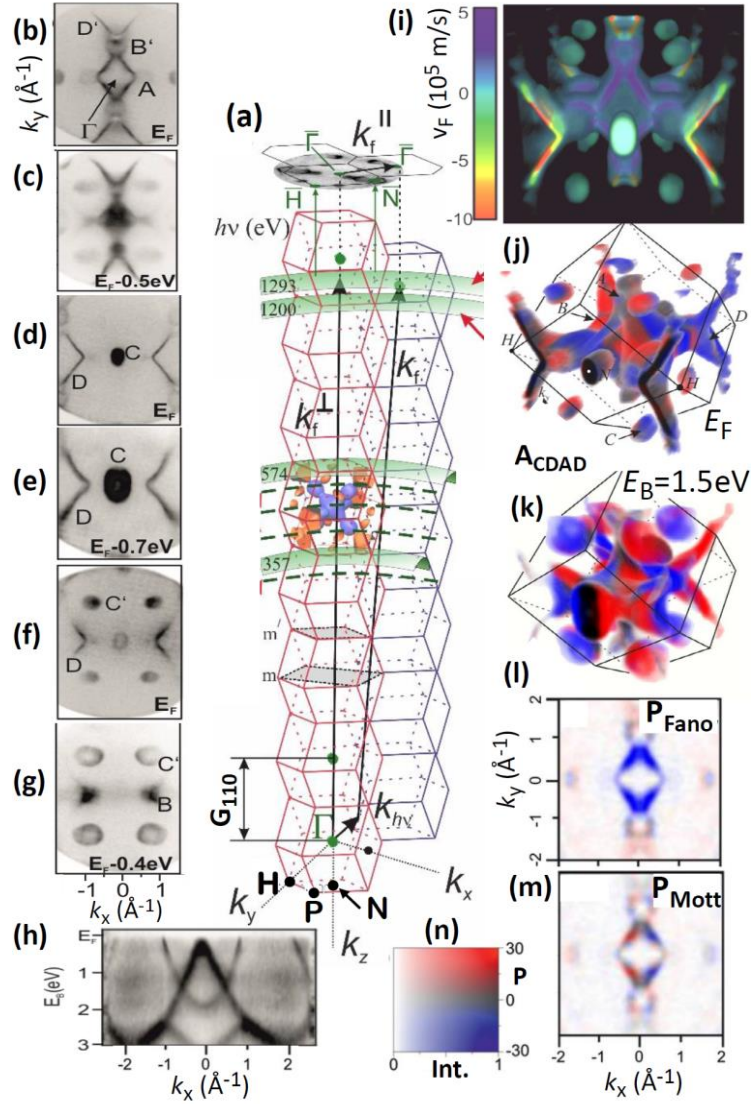

**FIG. 4**. Transition scheme for soft-X-ray photoemission from W(110) in a range of photon energies from 306 to 1293 eV and experimental results for intensity, circular dichroism and spin polarization. (a) Quantitative $k$-space description of transitions with final-state momentum vectors $k_f$ between 3.32 $G_{110}$ and 6.75 $G_{110}$; the photon momentum $k_{hv}$ is transferred to the photoelectron. Γ, H, N, P: high-symmetry points. (b-g) $k_x$-$k_y$ sections measured at different positions in k-space as marked in the transition scheme. Photon energies and binding energies stated in the panels, A-D mark different electron and hole pockets. (h) $E_B$-$k_x$ section showing the band dispersions. (i) Fermi surface extracted from a 4D array I($E_B$,**k**), concatenated from many 3D arrays. The colors show the Fermi-velocity distribution v($E_F$,**k**), see color bar [negative values denote hole conductivity]. (j) Fermi surface with overlaid circular dichroism asymmetry $A_{CDAD}$; red and blue mark positive and negative $A_{CDAD}$. (k) Same, but for the energy isosurface at $E_B$=1.5 eV. (l,m) Spin components $P_{Fano}(E_F,k_x,k_y)$ and $P_{Mott}(E_F,k_x,k_y)$, measured at hν= 460 eV; (n) color code for (j-m). All data recorded at beamline P04, PETRA III (DESY, Hamburg). (b-i) Reprinted with permission from Ref. 34, copyright 2017 Springer Nature; (j,k) reprinted with permission from Ref. 156 under a Creative Commons License; (l,m) reprinted with permission from Ref. 159, copyright 2019 IOP Publishing.

The 4$^{th}$ repeated BZ in Fig. 4(a) has been mapped in detail by recording ~20 individual 3D data sets with small increments of hν (i.e. different values of $k_z$)[34,156]. These sets are concatenated forming the array $I(E_B,\mathbf{k})$, which contains the full information on the electronic structure accessible by photoemission ('removal part' of the spectral function $\rho(E_B,\mathbf{k})$). From this array one can extract all *band dispersions* [example in Fig. 4(h)], all energy isosurfaces like the *Fermi surface* $I(E_F,\mathbf{k})$ with color-coded *Fermi velocity* $\mathbf{v}_F$ [Fig. 4(i)] and all other group velocities of the electrons $\mathbf{v}(E_B,\mathbf{k})$. The slopes of the bands have been extracted from many plots like Fig. 4(h) by numerical differentiation.

Such 4D k-space patterns $I(E_B,\mathbf{k})$ can be measured with hemispherical analyzers as well, either scanning the polar emission angle [157], or using a deflector in the entrance lens [158]. Specific features of the ToF-MM are the high acquisition speed, the linear k-scale and the large data arrays, comprising more than a full BZ and the entire d-band complex. Quantitative estimates of acquisition times will be given in the outlook in Sec. IV.

Measurements with circularly polarized X-rays at beamline P04 [140] gave access to further quantities: Fig. 4(j,k) *Circular dichroism asymmetry* $A_{CDAD}(E_B,\mathbf{k})$ and Fig. 4(l,m) *spin components* $P_{Fano}(E_F,k_x,k_y)$ and $P_{Mott}(E_F,k_x,k_y)$. The spin quantities have been measured at hν= 460eV using the upper ToF drift section in Fig. 1. Note the different symmetry properties of $P_{Fano}$ and $P_{Mott}$. Detailed analysis of CDAD and spin components using Feynman graphs revealed a previously-overlooked relation between these three quantities. The optical spin-orientation $P_{Fano}$ originates from the interaction with circularly polarized light (Fano effect) and the spin polarization component $P_{Mott}$ is induced by spin–orbit scattering (Mott scattering). The latter component is proportional to the product of the polarization stemming from the Fano effect and the CDAD asymmetry. Because both asymmetries can simultaneously change sign with the photon polarization, a finite spin polarization remains present even for excitation with linearly polarized light; for details, see Ref. 159.

All group velocities of the band electrons $\mathbf{v}(E_B,\mathbf{k})$ (oriented perpendicular to the corresponding isosurface) can be determined from the distance of adjacent constant-energy surfaces (e.g., $E_F$ and $E_F-dE$) using the gradient in k-space:

$$\mathbf{v}_F(\mathbf{k}) = \frac{1}{\hbar}\nabla_{\mathbf{k}}E(k)|_{E=E_F}. \qquad (5).$$

For tungsten, $\mathbf{v}_F$ varies from $10^5$ to $10^6$ m/s and $-10^5$ to $-2.5 \times 10^5$ m/s for electron and hole pockets, respectively.

The spin-integral measurements yielding Fermi surface, Fermi velocity and CDAD (via two measurements reversing the photon helicity for each photon energy) in Figs. 4(b-k) have been captured within ~5 hours at beamline P04 of the storage ring PETRA III at DESY in Hamburg. The 40-bunch mode of PETRA III corresponds to a pulse period of 192 ns that perfectly matches the count-rate capability (5 Mcps) of the delay-line detector (DLD). The spin measurements Figs. 4(l,m) required a total acquisition time of approx. 12 h.

***III.D. Hard X-ray ARPES: Temperature-dependent Fermi surface of the Kondo system YbRh$_2$Si$_2$ 3D bulk band mapping and hXPD***

The experimental results described below have been measured at beamline P22 [160] of the storage ring PETRA III. The photon band width at 5 keV using the Si(331) monochromator crystal is 155 meV, determining the energy resolution. The time-of-flight momentum microscope used for this experiment has been described in Ref. 149.

While offering the highest energy resolution, VUV excitation results in a pronounced surface sensitivity, where electronic properties could be altered by the broken symmetry at the surface. Therefore, bulk sensitive photoemission measurements using hard X-ray excitation provide an important complementary approach for understanding electronic properties of complex materials and buried interfaces[161, 162]. ARPES with hard X-ray excitation (HARPES) is challenging, because of the dropping cross sections and the increasing amount of photoelectrons that are scattered at phonons. The following example (from Ref. 163) shows that despite the scattering background, ToF momentum microscopy enables a determination of the band dispersion for complex compounds. The results are obtained for the model Kondo-Lattice system $YbRh_2Si_2$. Previous photoemission experiments on this material have focused on ARPES in the VUV regime, leaving open questions on the temperature dependence of bulk electronic states.

Before discussing the example, we address several peculiarities of HARPES: the role of the photon momentum, the recoil effect, and the nature of Laue and Kikuchi diffraction, all being negligible at small energies but substantial for hard X-ray photoemission. The first effect is related to the high *photon momentum*. Using eq. (2) we derive transversal and perpendicular components of the photon momentum as $\Delta k_y$ = 3.6 Å$^{-1}$ and $\Delta k_z$ = 1.3 Å$^{-1}$ for the highest photon energy used at P22 (h$\nu$ = 7.3 keV). This means that the final-state energy isosphere in *k*-space is shifted laterally by more than one Brillouin zone. In the momentum image, the Γ point of the valence band pattern (corresponding to the initial state with ***k***=(0,0,0)) is shifted away from the normal emission direction. The z-component of the shift (pointing towards the surface) must be accounted for in the tomographic-like *k*-space mapping as described in Fig. 4. $\Delta k_z$ shifts the isosphere to lower $k_z$ so that a different plane in the periodic BZs is mapped.

The *recoil effect* is related to the photo*electron* momentum. In the HAXPES regime the photoelectron momentum is an order of magnitude larger than the photon momentum; at $E_{final}$ = 7.3 keV the momentum is |***k***$_{final}$| = 43.8 Å$^{-1}$. In order to fulfil the total momentum balance in the photoemission process, this electron momentum is causing a recoil momentum transferred to the emitter atom. For HAXPES photoelectrons, the recoil momentum imparted to the atoms is no longer negligible. For light atoms it can be observed as a shift to lower kinetic energy in the 10 meV range. Since this shift depends on the atomic mass, recoil spectroscopy can yield information on the local environments of the emitter atoms in a material (for details, see Ref. 164). There are still open questions regarding the recoil effect for photoemission from itinerant band states, possibly resulting in recoil-free photoelectron emission[165, 166].

*Laue-type diffraction* of photoelectrons can be described in terms of Umklapp processes on the final-state energy isosphere in *k*-space. This effect is validated by many experimental examples and can be visualized in a graphical model[155]. Energy conservation demands that the final state stays on the isosphere (which is shifted by the photon momentum). The Umklapp changes the momentum by a reciprocal lattice vector ***G*** (Laue eq. $\Delta$***k*** = ***G***). This diffraction process retains the coherence and the entirety of all possible Umklapp processes constitutes the Pendry multiple-scattering final state of the outgoing photoelectron wave. The transfer of photon momentum, the recoil effect and Laue-type diffraction are *intrinsic* in the photoemission process.

The mechanism leading to *Kikuchi diffraction* is *extrinsic*, i.e. it does not happen at the emitter site. When travelling through the solid, the photoelectron can be scattered quasi-elastically at another atom. The scattering event destroys the coherence and extinguishes the initial momentum information. This incoherent contribution increases with increasing kinetic energy of the photoelectrons and increasing thermal disorder, i.e. with increasing temperature (governed by the Debye-Waller factor). This is the classical Kikuchi effect[167], except that the initial state is the photoelectron wave and not the primary electron beam of an SEM or TEM[168]. Kikuchi diffraction imprints a pronounced structure on this background, which can be eliminated by a multiplicative correction[169].

At energies of several keV the effective scattering volume comprises several $10^6$ atoms (in typical probe volumes of several $10^3$ nm$^3$). In this regime the diffraction process is well described by Bragg reflection at lattice planes[170], being operative at Kikuchi band edges. The widths of the bands are given by multiples of *reciprocal-space vectors*. The centre lines of the bands represent projections of *lattice planes (hkl)* and the crossing points of Kikuchi bands correspond to projections of *crystallographic directions [uvw].* Hence, Kikuchi patterns on the one hand provide a metric of *k*-space and on the other hand show the sample orientation. For strong core-level signals, such patterns are visible in real time with a frame rate of 1 Hz. Examples are shown in Figs. 5(a) und 6(a).

When the electron drops out of the 'Pendry wave field', its energy is almost retained except for the small atomic recoil energy. The classical elastic-scattering picture of a fast electron at an atom corresponds quantum-mechanically to a coherent multi-phonon excitation in a lattice[171,172]. The initial displacement of the scattering atom can be described as a local coherent phonon superposition, which stores the recoil energy. Elastic forces pull the displaced atom back towards its equilibrium position. Finally, the local lattice displacement evolves with time into a delocalized thermal excitation, characterized by the vibrations of the neighbouring atoms. In this way, the recoil energy ends up as thermal energy of the lattice. Calculations by Wang reveal that the cross sections for elastic scattering (diffraction) and inelastic scattering (also termed 'thermal diffuse scattering') are strongly different. Diffraction is peaked in forward direction (favouring small Bragg angles) and drops by orders of magnitude towards larger scattering angles. The phonon-mediated cross section varies only weakly with scattering angle, i.e., the inelastic channel appears rather isotropic (see Fig. 3 in Ref. 173; note the log scale). In the wave picture, the scattered electrons form a spherical wave centred at the scattering site. This wave is diffracted at the lattice as discussed in the early work of Kikuchi[167] and later Laue[174].

The heavy-fermion behavior in intermetallic compounds is characterized by a drastic increase of the effective mass of conduction electrons[175]. In YbRh$_2$Si$_2$ this behavior is caused by interactions between the strongly localized 4*f* electrons and itinerant electrons[176]. One expects that interactions between the local moments and the itinerant electrons leads to a transition from a small Fermi surface in a non-coherent regime at high temperatures to a large Fermi surface below a critical temperature[177,178]. Thus, the formation of correlated states changes the momentum dependence of electronic band structure. In a simple picture, the localized 4*f*-electrons involved in the hybridization are included in the Fermi surface encompassing all occupied states. If the magnetic moments related to the 4*f* electrons become unscreened, which may be provoked by lifting the coupling of conduction electrons with the local moments above a critical temperature, the underlying 4*f*-electrons become localized again and are thus excluded from the Fermi surface. The size of the Fermi surface is therefore increased in the Kondo lattice state in comparison to the uncorrelated state.

An interesting question is at which temperature the change of the band dispersion occurs. The periodic Anderson model predicts the loss of coherence near the Kondo temperature $T_K$=25 K[179]. In contrast, dynamical mean field theory calculations result in a continuous change of the Fermi surface far above $T_K$[180]. VUV -ARPES[181,182] results in a temperature-independent Fermi surface between 1 and 100 K[183,184]. The formation of a 4*f*-derived flat band has been observed above 120 K by scanning tunneling spectroscopy[185] and ARPES[186].

A particularly intriguing advantage of momentum microscopy with a large k-field-of-view is the combination of valence-band mapping with high-resolution *hard X-ray photoelectron diffraction* (hXPD). hXPD is sensitive on the geometric structure. In the hard-X-ray regime the short photoelectron wavelength reveals tiny phase differences and turns hXPD into a very sensitive structural tool[22]. Calculations reveal detectable changes of the lattice constant in the 1% range by evaluating the fine structure near the zone axis[22]. This suggests possible applications of hXPD for element-resolved strain analysis in materials [187, 188]. Valence-band patterns are recorded at identical settings of the microscope. This allows the direct comparison of changes in the geometric and electronic structures. Static and dynamic strain manipulation emerges as a new avenue to tune and shape a material's electronic properties[189,190].

Fig. 5(a) depicts the hXPD pattern obtained at the Yb 4$f$ core level. Prominent features are the two diagonal (110)–type Kikuchi bands, one being marked by lines on both sides. Their widths are determined by the reciprocal lattice vectors $G_{110}$ and $G_{1-10}$ (marked width 3.1 Å$^{-1}$). The rich structure of the pattern is caused by the position of the Yb atoms within the crystal structure. Details of this structure are revealed by comparison with simulated patterns.

A practical aspect is the possibility to determine the orientation of the sample using pronounced Kikuchi bands. Thanks to the high intensity of the core-level signals, such hXPD-patterns are visible with good contrast in real-time with frame rates of 1 Hz. The straight lines can then be used to adjust the electron optics avoiding image-field distortions at large $k$-field diameters. The size of the projected BZ is immediately visible in the Kikuchi patterns, providing a metric in $k$-space.

The large momentum field of view of 12 Å$^{-1}$ allows the simultaneous acquisition of five adjacent Brillouin zones. This large field of view can be exploited to measure the dispersion of valence bands in 3-dimensional k-space without changing the photon energy. As sketched in Fig. 5(c), the perpendicular momentum component along $k_z$ for the probed intensity decreases with increasing parallel momentum due to the energy conservation law, forcing all final states on an energy isosphere with radius $k_f$ [see Eq. (1)]. This fact allows simultaneous data acquisition as a function of perpendicular momentum $k_z$, covering a range of $\Delta k_z = 0.5\ G_{001}$ in one experimental run.

The photon energy set to $h\nu$ = 5297 eV results in $k_z$ = 29 $G_{001}$[163]. As discussed above, the substantial shift by the photon momentum needs to be taken into account. Adding the corresponding reciprocal lattice vector, data processing transforms the measured 3-dimensional array $I(E_B,k_x,k_y)$ into the 4-dimensional data array $I(E_B,k_x,k_y,k_z)$. To visualize the data, surfaces of constant energy in three-dimensional $k$-space result from cuts of the four-dimensional array as shown in Fig. 5(d) for 4 different binding energies. For $E_B$=0 the doughnut (D) is centered at the Z point. The matrix-element-induced intensity variations fractionate the jungle gym (J) surface centered at the Γ point. Figs. 5(e-g) depict the constant-energy surfaces at the indicated binding energies. Due to the hole-like dispersion of bands D and J the spectral densities retract towards the Brillouin zone edges. Lower-lying bands appear at higher binding energies near the Z and X points [Fig. 5(g)].

The dispersion of band D (doughnut) at low temperature shown in Figs. 5(e-g) reveals significant differences as compared to corresponding data measured at room temperature [Figs. 5(i-k)]. This difference also shows up in the cuts revealing the energy dispersions $E_B(k)$ along the Z-Σ high symmetry direction [Figs. 5(h,l)]. Note that for this representation the intensity for each binding energy has been normalized to the Brillouin-zone averaged intensity to suppress the high momentum-independent intensity of the 4$f$ states. The normalization emphasizes the dispersion of electronic bands on the expense of suppressing the signal from the non-dispersive part of the localized 4$f$-states. The solid blue line in Fig. 5(h) indicates the position and dispersion of band D at 20 K. The corresponding band dispersion measured at 300 K shown in Fig. 5(l) (red line) shows that band D is shifted to higher binding energies. The energy shift amounts to 100 meV.

Similar data sets have been measured for several temperatures between 20 and 300 K [Fig. 5(m)]. The distance $K_{Z\Sigma}$ in k-space from the Z point to the position of band D along ZΣ illustrates the temperature dependence of the band shift. At 20 K this distance is significantly larger than at 300 K. The change occurs between 100 K and 200 K. The characteristic temperature $T^*$ for the change of the electronic structure is thus considerably larger than the Kondo temperature of this compound, i.e. $T^* > 4T_K$.

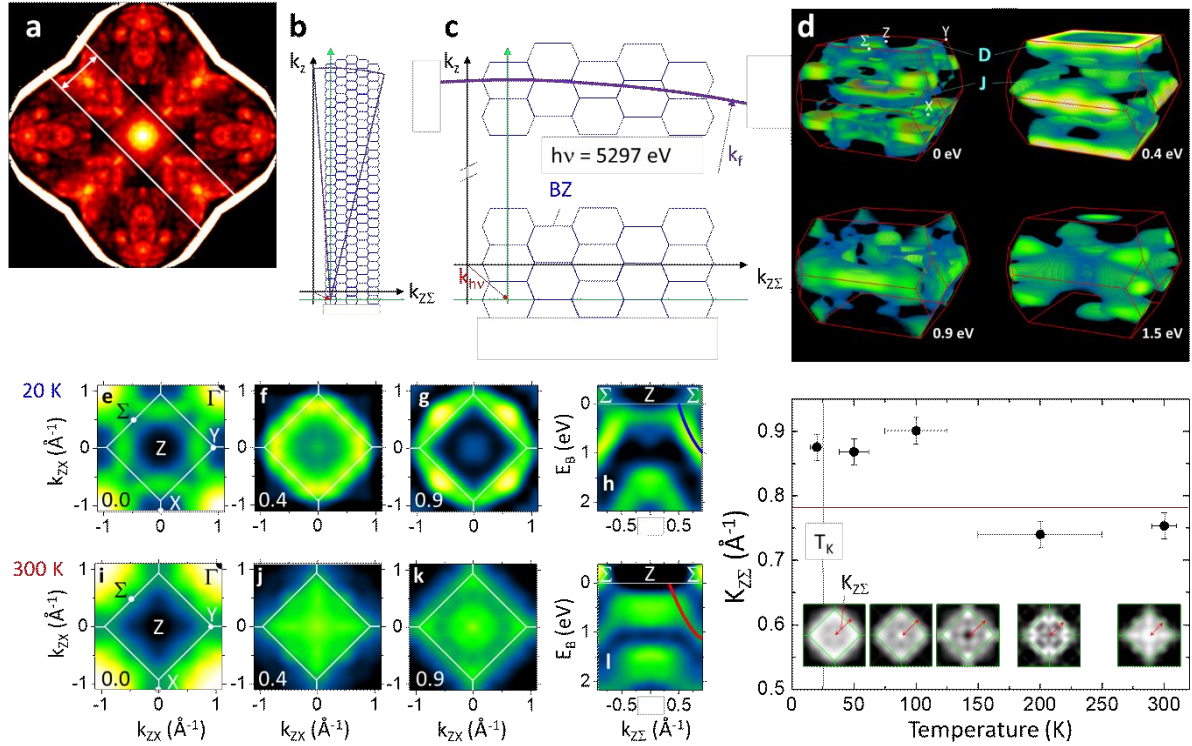

**FIG. 5.** (a) Hard-X-ray photoelectron diffraction (hXPD) patterns of the Yb 4f level in YbRh$_2$Si$_2$ at a photon energy of 5300 eV. The width of the prominent Kikuchi band (double arrow in (a)) is given by $G_{110}$ =3.1 Å$^{-1}$. (b,c) Momentum-space description of a photo-transition from the valence band. ($k_x$,$k_z$)-scheme describing direct interband transitions into a quasi-free-electron-like final state band at a photon energy of 5297 eV. The horizontal axis corresponds to the ZX direction. The plots are to scale for the YbRh$_2$Si$_2$ structure. The final-state momentum $k_f$ (radius of the final-state sphere) is approximately independent on $E_B$ (for high photon energy). The center of the sphere is displaced from the origin k = (0,0,0) by the vector of the photon momentum $k_{hv}$. Brillouin zones (BZ) are depicted in the periodic zone scheme. (d) Three-dimensional representation of the measured spectral density of states of YbRh$_2$Si$_2$ at four indicated binding energies, measured at a photon energy of 5297 eV at a temperature of 20 K. (e-g) Constant-energy maps at the indicated binding energies of the spectral density of electronic states in the Z–Y–Σ plane measured at 20 K. (h) Energy dispersion along Z–Σ (dark: lowest and yellow: highest intensity). (i-l) Analogous data measured at 300 K. Blue and red lines in (h) and (l) indicate the different band dispersion of band D. (m) Temperature dependence of the radius of band D along the Z–Σ direction at a binding energy of 0.4 eV. The solid red line marks the Z-Σ distance. The insets show constant energy maps for the corresponding temperatures (white means high intensity). (b-m) Adapted from Ref. 163 with permission; copyright 2021 Institute of Physics (UK).

While this observation is inconsistent with the prediction of the periodic Anderson model, it confirms previously observed changes in core level spectra, resonant X-ray electron scattering [184] and scanning tunneling spectroscopy [186]. The temperature-dependent changes of the optical conductivity spectra [191] also indicate changes of electronic states at much higher temperature than $T_K$. The results also agree with recent time-resolved ARPES experiments, indicating a decrease of the f–d hybridization with increasing electron temperature around 250 K [192].

This example demonstrates the applicability of angle-resolved photoemission spectroscopy in the hard X-ray regime (HARPES) for complex materials in a larger temperature regime. Despite the large background due to thermal diffuse scattering of photoelectrons with high kinetic energy, the high-efficiency of data recording using ToF MM enables the extraction of detailed bulk information. A few hours of data acquisition reduce the statistical noise to such a level that background subtraction reveals the information encoded in the small part of photoelectrons that are not scattered. Further examples show that it is possible to determine the band dispersion through capped surfaces. This is an important advantage because it eases sample handling and allows sample preparation techniques such as application of large magnetic fields[193] and lithography techniques that otherwise are prohibitive for photoemission spectroscopy.

### III.E. Spin-resolved HAXPES: Is magnetite indeed a half-metallic ferromagnet?

As has been discussed in the previous chapter, hard X-rays enable a new avenue towards band mapping of capped complex bulk materials, buried layers or interfaces in thin-film architectures as well as in *in-situ* and *in-operando* devices. Spin-resolved HAXPES substantially broadens the knowledge revealed by photo-emitted electrons. For example, one may envision measuring the spin polarization at buried interface states [194] and even for *in-operando* devices, where external electrical fields modulate spin-momentum locked states by spin-orbit enhanced interactions [195, 196].

The scattering of photoelectrons with phonons may pose an obstacle for spin-resolved HAXPES. The scattering event of the photoelectron may not only alter its momentum but the spin orientation, too. The fraction of photoelectrons without scattering, described by the Debye Waller factor, exponentially decreases with increasing kinetic energy. This is in particular serious for ferromagnetic materials typically consisting of light elements and having a low Debye temperature. For materials with heavy elements additional effects have to be considered. As has been discussed for the case of soft X-ray photoelectron spectroscopy, spin-orbit interaction in the final state contributes to the spin-polarization of photoelectrons [Fig. 4(l,m) and Ref. 159].

The investigation of ferromagnetic samples provides a straightforward approach to test the applicability of spin resolved HAXPES[197,198]. Here, the spin polarization of photo-emitted electrons is determined from measurements of two remnant magnetization states with opposite direction. This procedure allows to eliminate spin-orbit induced additional spin-polarization effects[159]. The question whether phonon scattering reduces the observable polarization remains. To answer this question, we discuss in the following spin-resolved HAXPES data obtained for the case of magnetite[199], a material with predicted half-metallic properties.

Ferrimagnetic magnetite ($\alpha$-Fe$_3$O$_4$), historically the oldest ferromagnetic material applied in technology, is still interesting for modern research because of potential applications in spintronics[200]. Magnetite has a high Curie temperature of 858 K[201] and according to a simple model[202], the itinerant conduction states are pure uncompensated minority states. Magnetite is thus a candidate material for a half-metallic ferromagnet. Density functional theory has confirmed this simple model[203,204]. In contrast, a single ion model considering the localized character of electronic states in oxides, concludes that the spin polarization has a considerably lower value of P=-2/3[205]. Experimental values from spin-resolved photoemission studies obtained with excitation energies between 10 eV and several 100 eV range from $P$ = +0.3 to -0.8 [204,206,207,208,209,210,211,212,213, 214, 215,216].

This discrepancy may be explained by the low probing depth of a few atomic layers. The inverse-spinel structure of magnetite is vulnerable at the surface because all surface planes are charged[217]. To avoid a large polarization the crystal surface tends to reconstructions[218,219,220]. Moreover, the preparation of clean magnetite surfaces appears to be difficult to achieve[218]. Spin-resolved HAXPES is attractive for solving the question of bulk spin polarization because its large probing depth promises true bulk information and thus a better reproducibility of the results.

The spin resolved HAXPES results that we exemplarily discuss in the following used the high-energy version[149] of the setup described in Fig. 1 that operates in the hard X-ray regime and that was installed at beamline P22[160] at PETRA III (Desy, Hamburg). The instrument allows for simultaneous energy-, spin- and momentum-resolved mapping of the photoelectrons with unprecedented efficiency. The sample is an epitaxial $\alpha$-Fe$_3$O$_4$(111) film of 30 nm thickness grown on ZnO(0001) by molecular beam epitaxy[221].

Hard X-ray photoelectron diffraction (hXPD) patterns [Fig. 6(a)] measured in the same run confirm the structural quality of the investigated film. hXPD gives detailed information about the geometric structure in the identical sample volume probed for the spin-polarization[222]. Figure 6(a) displays the measured hXPD patterns for the Fe 3p and O 2p core levels in comparison to simulated patterns based on the many-beam dynamical theory of electron diffraction[223]. The good agreement of theory and experiment confirms the good structural quality of the sample and supports the bulk sensitivity of the photoemission measurement at $E_{kin}$= 5.15 keV.

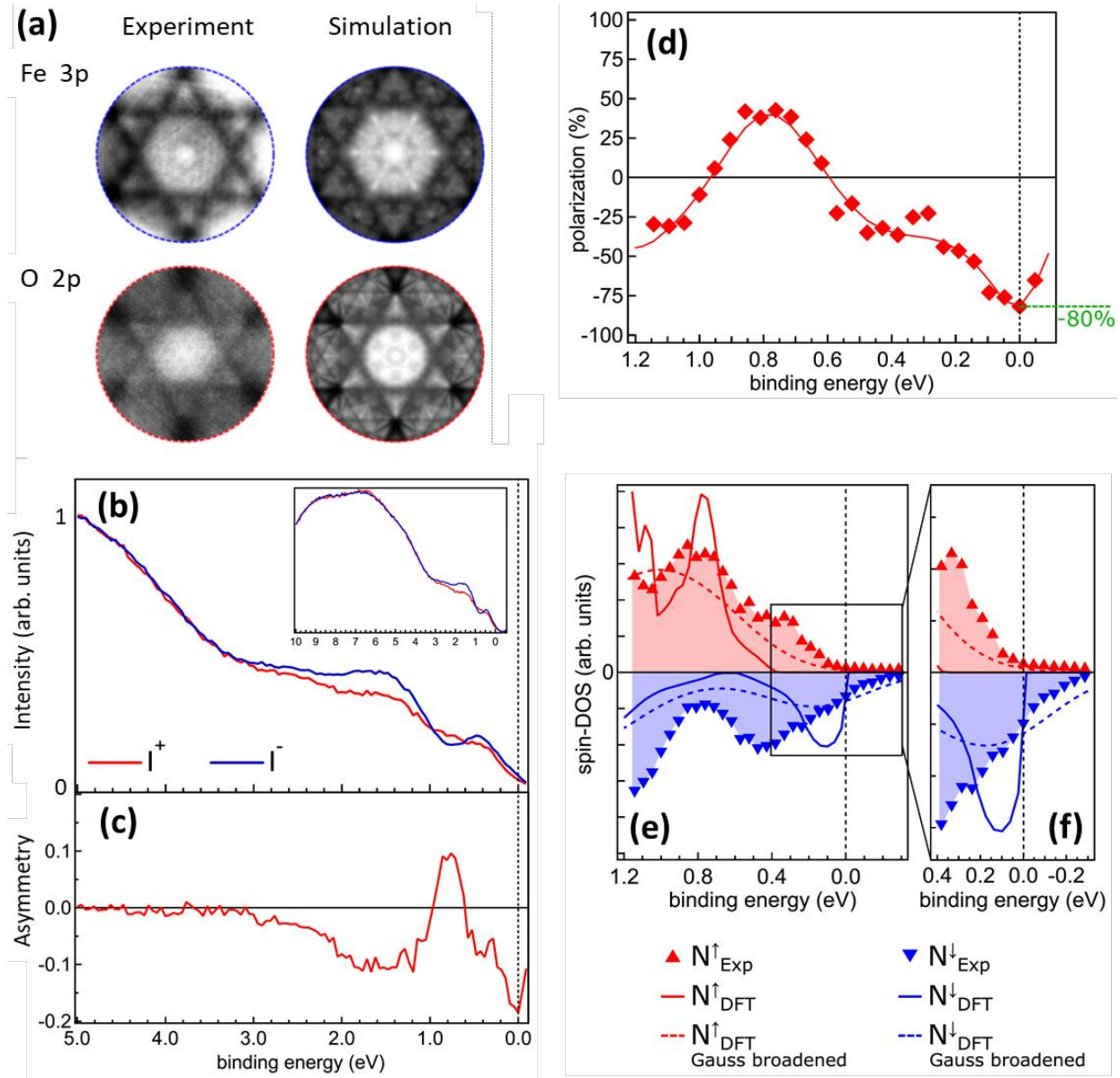

**FIG. 6.** (a) hXPD patterns of the Fe 3p ($E_{kin}$ = 5150 eV) and O 2p ($E_{kin}$ = 4993 eV) core levels in magnetite: experimental data on the left and calculated diffraction patterns on the right are in good agreement. The hXPD patterns cover an angular range of +/- 9°. (b) Valence band spectra measured with a photon energy of 5.0 keV and at 30 K for two opposite in-plane magnetizations denoted by I$^+$ and I$^-$. (c) Energy dependent asymmetry A, calculated from the data shown in (b). (d) Spin polarization as calculated from (c) using the calibrated energy-dependent Sherman function $S(E_B)$. The spin-polarization is negative at $E_F$ = 0 eV with a value of -0.8. (e,f) Majority and minority spin-DOS (red and blue triangles, respectively), calculated from the results shown in (b) and (c). Results from DFT calculations (solid lines, Ref. 226) are shown for comparison considering the instrumental energy resolution (dashed lines). Adapted from Ref. 199 with permission; copyright 2021 American Physical Society.

The valence band was measured with an excitation energy of 5.0 keV, leading to a photoelectron inelastic mean free path of 70 Å [224] and an energy resolution of 620 meV using a Si(111) monochromator crystal pair for maximum photon flux. Due to the weak dispersion of the Fe 3*d* bands their momentum dependence cannot be resolved [225]. Therefore, the spin-resolved data are integrated over an area corresponding to more than a full Brillouin zone. To retrieve the spin information, photoelectron spectra are recorded for two opposite magnetization directions after reflection at the spin-filter crystal as shown in Fig. 6(b). The spin-dependent reflectivity at the surface of an iridium single crystal at kinetic energies near 12.5 eV [101] is exploited to calculate the asymmetry from the two spectra [Figure 6(c)]. The Sherman function $S$I determining the spin sensitivity has been calibrated prior to the experiment. Near 12.5 eI($E$) is approximately 0.22. The spin-polarization determined from the asymmetry by dividing Ih $S(E)$ [Figure 6(d)] shows a maximum negative value at the Fermi level of -80%. Because of the monotonously decreasing spin polarization with increasing binding energy followed by a sign change at $E_B$ = 0.6 eV, the Fermi level value represents a lower boundary of the true bulk spin polarization. A better energy resolution would lead to an even higher (negative) value closer to -1. Thus, the experimental result is consistent with half-metallic behavior.

Assuming a constant value for the photoemission matrix elements, majority and minority spin-densities are calculated from intensity and spin-polarization [see Figure 6(e,f)]. The spin-resolved density-of-states provides a good starting point for the comparison with theory. The DFT calculations[226] are shown with and without considering the experimental broadening. The experimental spin-densities overall resemble the theoretical ones. Remaining differences are attributed to electron correlations of the itinerant valence states and to their coupling to the lattice degrees of freedom.

This example shows that the new generation of spin-resolved time-of-flight photoemission spectrometers enable the determination of the bulk spin polarization of magnetite by strongly enhancing the probing depth with hard X-ray excitation. With such a setup, it has become possible to overcome the notorious problem of in-situ surface preparation and often ill-defined surfaces. Furthermore, the large value of the spin-polarization reported here confirms that phonon scattering processes of photo-emitted electrons do not significantly decrease their spin polarization.

### *III.F. Free-Electron-Laser and HHG ARPES: Femtosecond dynamics of valence and core electrons*

Time-resolved photoemission is the youngest member of the large photoemission family, closely connected to the development of fs-pulsed photon sources. While an increasing number of laboratory-based HHG sources for photoemission are available or being built, the FEL-based photoemission is gaining visibility, too. Here we will have a short view on the present status and future potential of the ToF-MM technique in the field of fs pump-probe experiments.

The discussed experiments using FEL radiation have been performed using the HEXTOF (ToF-MM with low-energy optics) at FLASH in cooperation of the FLASH team with several user groups. The energy range of the PG2 beamline (25 – 830 eV) provides access to both the valence band as well as core levels. Figures 7(a-d) show measurements on WSe$_2$ recorded using the HEXTOF (adapted from Refs. 42,43). The wide-range ToF spectrum (a) shows the valence band (VB) and the outermost core levels (W 4*f* and Se 3*d*, at $E_B$= 33 and 55 eV, respectively), whose responses to the photoexcitation have been studied for hν$_{pump}$= 1.6 eV and hν$_{probe}$= 36.5 eV for the valence range and hν$_{probe}$= 109.5 eV (3$^{rd}$ harmonic) for the core levels. hν$_{pump}$ is resonant with the A-exciton transition in WSe$_2$, i.e., the pump photons populate the K and K' valleys of the conduction band. Figs. 7(b) and (c) show two snapshots from the $I(E_{final},k_x,k_y,\tau)$ data array, (b) before the pump pulse ('probe only') and (c) after the pump pulse (τ > 0). The population of the K and K' valleys is clearly visible in (c). Tracking the temporal evolution of the population reveals a fast-decay component with a time constant of ∼80 fs. From K and K' the electrons are rapidly scattered into the lower-energy Σ and Σ' valleys, where the population reaches its maximum with a delay of ∼60 fs compared to the K-valley signal. The Σ / Σ' population shows a multi-exponential relaxation dynamics with a faster time constant of ∼160 fs (for details, see Fig. 5 in Ref. 42).

A special advantage of the HEXTOF instrument is its capability of recording valence-band and core-level dynamics with identical settings; only the probe photon energy and the sample-bias voltage are changed. Thus, the dynamics of valence electrons could be directly compared with the concurrent changes induced in the spectral function of the core level. Fig. 7(d) shows a k-integral representation of the W $4f_{5/2}$ core-level signal as function of delay time τ between -0.5 and +2.4 ps. Here the energy scale is referenced to the binding energy of W $4f_{5/2}$ in the unpumped case (τ<0). Upon arrival of the pump pulse at τ=0 the linewidth and position change. Close inspection of the relative dynamics of the conduction-band signal [(c) and its temporal evolution] and core level signals [(d) and the equivalent for Se $3d$ [42]] clearly show a delayed core-hole renormalization due to screening by excited quasi-free carriers resulting from an excitonic Mott transition. These findings establish time-resolved core-level photoelectron spectroscopy as a sensitive probe of subtle electronic many-body interactions and ultrafast electronic phase transitions. Dendzik et al. [43] created the term 'core-*cum*-conduction' photoemission spectroscopy for this kind of experiment.

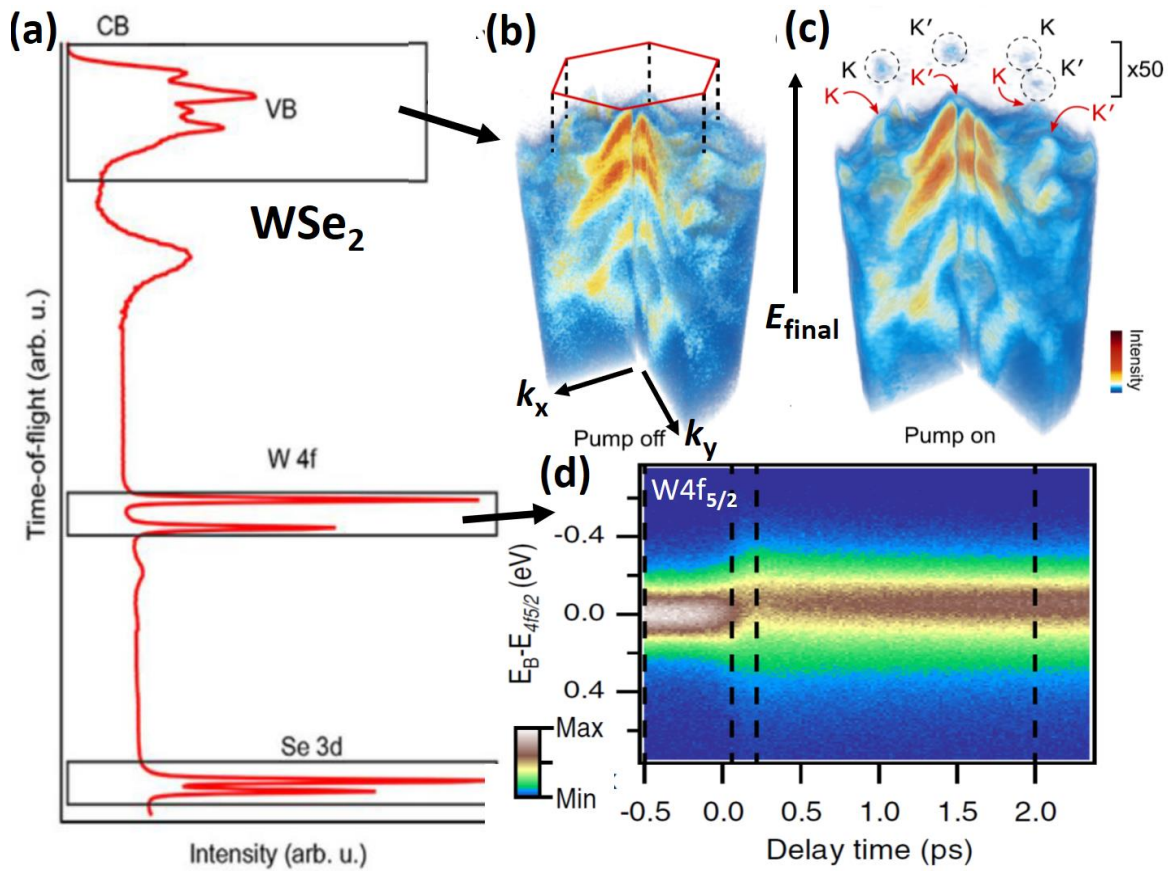

**FIG. 7.** Pump-and-probe time-resolved photoemission results for WSe$_2$, recorded at the PG2 beamline of the free-electron laser FLASH at DESY (Hamburg). (a) Time-of-flight spectrum showing the valence range and W $4f$ / Se $3d$ core levels. (b,c) 3D representations of the measured $I(k_x,k_y,E_{final},\tau)$ data array of the valence-band range recorded with hν$_{probe}$ = 36.5eV (b) without pump pulse and (c) after a pump pulse (hν$_{pump}$=1.6eV, s-polarized), populating the K and K' valleys of the conduction band. (d) k-integral measurement of the W $4f_{5/2}$ core-level signal as function of delay time τ. Intensity scales, see color bars in (c,d).
(a-c) Reprinted with permission from Ref. 42, copyright 2020 American Institute of Physics; (d) reprinted with permission from Ref. 43, under a Creative Commons License.

*Time-resolved ToF-MM at HHG sources* is facing exciting prospects for the near future. On the experimental side, HHG sources with advanced performance are coming up, offering pulse rates towards 1 MHz or more and photon energies beyond 100 eV. A particularly exciting development is the possibility to work in a regime without significant space-charge effects using an intra-cavity-enhanced HHG source, running at 60-80 MHz pulse rate[227]. The extremely high count rates (comparable to the conditions discussed in Sec. III.A) allow exploiting the *small-area mode* of the ToF-MM. Regions of interest down to the sub-micrometer range[35,88,89,90] can be defined by a small field aperture in the first Gaussian image [Fig. 1(a)]. In the PEEM-imaging mode, this aperture allows precise positioning of the analyzed area on heterogeneous samples, independent on the size of the photon footprint (e.g., Fig. 1(c) in Ref. 89 and Ref. 90). The Stony Brook instrument (ToF-MM at a 61 MHz intra-cavity-enhanced HHG source[228], providing tunable fs radiation in the range of $h\nu = 10 - 40$ eV) is setting a new benchmark of performance in ultrafast visible-to-near-IR pump / HHG-probe experiments. First experiments, tracking the hot-electron relaxation dynamics in HOPG[229] and excited-state dynamics in the bulk conduction band of the prototypical transition metal dichalcogenide $WS_2$ [230] provide a flavor of the data quality that resembles previous static experiments. A stronghold of this approach is the perturbative limit, where space-charge effects are largely negligible and the observed dynamics is dominated by the intrinsic quasiparticle behavior.

A similar leap in ToF-MM performance is expected with the advent of 'quasi-continuous' FEL radiation. LCLS-II, for instance, will deliver pulse rates of up to 1 MHz [46], as opposed to 5000 pulses per second, presently possible at FLASH (e.g., experiment in Fig. 7). The data-acquisition rate will be further enhanced by new generations of detectors with high multi-hit capability[94-97].

Given such an advanced technological basis, numerous exciting science cases can be tackled. A prominent example is the detailed analysis of the full wealth of momentum- and energy-dependent population dynamics accessible by tr-MM (as for tr-ARPES in general). The role of the electron-phonon interaction is very complex as discussed in detail by Sentef, Devereaux et al.[52,53]. In many systems the hot electrons initially lose energy via transfer to strongly coupled optical phonons. Only when the optical phonon bath has reached the hot-electron temperature, acoustic phonons are created, which happens on a longer intrinsic time scale. Yet, *e-e* energy dissipation channels can also be significant. Confirming a decade-old prediction, Curcio et al.[231] found that ultrafast line-width broadening in the C 1s core level of graphene is caused by an exchange of energy and momentum between the photo-emitted core electron and the hot-electron system, rather than by vibrational excitations. A line-shape analysis accounting for the presence of the hot-electrons bath directly yields the transient electronic temperature of the system. Such experiments will be complemented by tr-XPD, providing valuable structural information, simultaneously to electronic structure mapping via tr-ARPES. A first pilot experiment using ToF-MM recently succeeded in observing coherent phonon excitation in $Bi_2Se_3$ [232].

We come back to the recent achievements of *wavefunction retrieval* and *dynamic orbital tomography* as mentioned at the end of Sec. II.A. Multidimensional MM data alongside with progress in theoretical understanding widen the information content towards what we termed 'complete experiment'. The Bloch wavefunctions themselves can be reconstructed,[85-90,233] including fundamental properties like exciton binding energy and exciton-lattice coupling. Beyond that, *polarization-modulated photoemission*[85] and the phase-sensitivity of CDAD[234] enable tracking the energy- and momentum-resolved amplitude and phase of the photoelectron wave. fs time resolution gives access to the dynamics on its intrinsic time scale. Probing the early-stage exciton dynamics in $WS_2$ revealed the formation of a momentum-forbidden dark KΣ-exciton a few tens of femtoseconds after optical excitation[87].

For a bilayer pentacene film on Ag(110), optical-laser-pump and FEL-probe experiments using ToF-MM established *time-resolved molecular orbital mapping*. It gives access to fundamental processes of excited wave packets and charge-transfer dynamics on femtosecond time scales. Time-dependent k-maps of the molecular valence states can be directly related to their molecular initial states. A change of the molecular orbital shape is identified as a structural and position rearrangement of molecules bound to the substrate during interfacial charge transfer[80]. For perylene-tetracarboxylic-dianhydride (PTCDA) on oxidized Cu, the full k-space distribution of transiently excited electrons connects their excited-state dynamics to real-space excitation pathways[81]. In molecules this distribution is closely linked to orbital shapes; hence, such experiments will offer the observation of ultrafast electron motion in time and space. For the copper(II) phthalocyanine (CuPc)-1T-TiSe$_2$ interface, the spectral evolution of the core levels yields important insight into the many-body response that follows the hot-carrier transfer in the valence region. The momentum distribution of the core electrons reveals a structural rearrangement as a result of interfacial hot-carrier transfer between valence orbitals and substrate bands[81,84]. These examples shine first light on the potential of *quasi-simultaneous valence and core level k-microscopy combined with photoelectron diffraction*, made feasible by the high photon energy of the FEL probe.

Concluding this section, we note that this review summarizes only experiments using the ToF-MM technique. Orbital tomography as well as photoelectron diffraction data can be recorded with excellent quality using hemispherical analyzers[235,236]. The future will show whether parallel 3D data acquisition in ToF-MMs can develop its full potential in time-resolved measurements. For the highly ambitious fs tr-ARPES experiments it is worthwhile elucidating strengths and weaknesses of different methods in detail. The space-charge suppression mode with retarding front lens is not yet fully explored. The proof-of-principle experiment[112] and first application[44] employed a standard ToF-MM with non-optimized front lens. A key component of such experiments is the detector, since non-optimized detector configurations (like the missing high-pass filter in the MM-experiment discussed in Ref. 237) can strongly reduce the overall performance.

## IV. CONCLUSIONS AND OUTLOOK

Looking back on a century of exciting photoemission experiments, launched by the earliest work on XPS[238], we are presently facing a quantum leap in advancement, largely driven by the advent of photon sources with ultimate performance. Synchrotron sources reach physical limits of emittance and brightness. Free-electron lasers and the upcoming attosecond few-cycle light sources can provide extreme single-pulse energies. Space-charge effects pose a limit to the number of usable photoelectrons per photon pulse. High-repetition-rate FELs like FLASH at DESY (Hamburg) or LCLS-II (Stanford) are thus of utmost importance for tr-ARPES and tr-XPD experiments.

This progress on the source side comes along with progress on the detector side. All high-performance photon sources have well-defined time structure, hence ToF techniques are rapidly gaining importance. Here we discussed the ToF-MM concept, combining parallel recording of many kinetic energies (encoded in the time-of-flight) with full-field ($k_x,k_y$) momentum imaging. The potential of the new method in a wide range of excitation energies from the UV to hard X-rays is illustrated by a number of examples. The examples demonstrate the benefit of highly-efficient 3D recording to spin- and time-resolved measurements and experiments using hard X-rays. Such experiments suffer from the low detection efficiency of spin filters, low repetition rates of present FELs, and low photoemission cross sections alongside with strong phonon scattering in the HAXPES range.

*Multichannel spin filters* act like spin-selective mirrors for full *k* images with large effective figure-of-merit. In a ToF-MM, full-field *k* imaging and parallel energy recording are retained in the spin-resolving ToF branch. Hence, spin-filtered 3D data arrays $P(E_{kin},k_x,k_y)$ are captured simultaneously. Suga and Tusche show a comparison of many types of spin detectors (see Fig. 28 in Ref. 100) and estimate effective figures-of-merit FoM$_{eff}$ in the range $10^2$-$10^4$ for such 3D spin filters. For the example in Sec. III.A. we derive FoM$_{eff}$ ≈ 400. This high FoM$_{eff}$ directly shows up in the count rates.

Employing the 4$^{th}$ harmonic of a Ti:sapphire oscillator (hν= 6eV; pulse rate 80MHz), count rates of almost $10^6$ counts per second have been achieved in the spin-detector branch of the ToF-MM (Fig. 1). Such count rates facilitated rapid mapping of the full spin-resolved band structure of several reactive Heusler compounds (III.A). First spin-ToF-MM experiments have been successful in the soft- and hard-X-ray range at the storage ring PETRA III (Hamburg). The large information depth for hard X-rays (hν= 5keV) has been exploited for capturing the true bulk spin polarization in magnetite (III.E), proving the half-metallic ferromagnetic nature of $Fe_3O_4$.

The high-energy ToF-MM can detect kinetic energies up to >7 keV and record k-field diameters up to >20 Å$^{-1}$. It thus captures an area in k-space an order of magnitude larger than the low-energy variant. This large k-acceptance efficiently counteracts the drop in cross sections in the HAXPES range. Beyond this intensity aspect, it opens new opportunities for new types of experiments. For such large k-fields-of-view the curved final-state energy isosphere intersects adjacent BZs at different $k_z$. This fact can be exploited for mapping the full 4D spectral function ρ($E_B$,**k**) *without the need to vary the photon energy*. This ultrafast recording scheme allowed us to observe the temperature dependence of the Fermi surface and band dispersion for the Kondo system YbRh$_2$Si$_2$ (III.D).

The large k-field also enables a new, highly effective approach to *X-ray photoelectron diffraction (XPD)*. Full-field-imaging of diffractograms reveals a complex pattern of intersecting Kikuchi bands. High angular resolution (typically 0.03°) combined with good energy resolution (given by the photon bandwidth) exploit the capabilities of the ToF-MM in an ideal way. Imaging XPD is particularly effective in the hard-X-ray range, where several $10^6$ atoms contribute to the multiple-scattering final state. In all cases studied so far (graphite, Si, Ge, Re, Mo, W, GaMnAs, TiTe$_2$, MoTe$_2$, NbSe$_2$, Fe$_3$O$_4$, SrTiO$_3$, YbRh$_2$Si$_2$, EuPd$_2$Si$_2$) hXPD captured patterns with unprecedented richness in details; an overview is given in Ref. 239. Next to a detailed structural analysis, such diffractograms have practical aspects: They provide a metric in k-space and show distinguished directions in real space. The widths of Kikuchi bands correspond to multiples of reciprocal lattice vectors. The center lines and crossing points of bands correspond to projected lattice planes and high-symmetry directions of the crystal. For intense core levels, diffractograms can be observed in real time with a frame rate of 1Hz, enabling rapid alignment of sample and microscope settings, despite the small cross sections in the hard-X-ray range. Combining XPD and valence-band mapping in a single experiment gives access to the interplay of geometric and electronic structure[23]; examples are shown in Figs. 5(a) and 6(a).

As an *outlook*, we will give a tentative answer to the question posed at the end of Section II.A, whether the concept of a 'complete' experiment as introduced for free atoms and molecules[75,76] can be transferred to solid-state photoemission. The estimation considers the full analysis (intensity, dichroism and spin) of a pump-pulse-induced phase transition probed with fs X-ray pulses. In the general case it is not *a priori* known, where in parameter space the interesting physics is going to happen. Section III.D shows an example, where the formation of the many-body state in a Kondo system comes with significant changes of the dispersion of certain bands and shape of the Fermi surface in a static experiment [($E_B$,**k**,T$_l$) parameter space].

Experimental Fermi surfaces [like those in Figs. 4(i) and 5(d)] are projected from 4D arrays covering a bit more than a full BZ in k-space. The $k_z$-dependence is tracked by recording typically 20 ($E_B$,$k_x$,$k_y$) stacks with ≥$10^4$ pixels in each of the ~100 energy slices. Concatenation yields an array with about $10^7$ 4D voxels for the static intensity. Dynamic measurements track typical time intervals of several ps with probe pulses in the 50-150 fs range, resolving ~100 delay-time values. The final 5D ($E_B$,**k**,τ) array thus consists of ~$10^9$ voxels. Reasonable information requires on average at least 10 counts per voxel, i.e. a total number of $10^{10}$ counts. High-repetition-rate fs laboratory sources can deliver pulse rates in the 1 MHz region[240,241] (intra-cavity enhanced HHG sources even much more[227,228]) and future FELs will also reach pulse rates of a MHz[46]. In the present experiments at FLASH, the number of photon pulses per second is small (up to 5000); hence these experiments have to improve counting statistics via data binning at the expense of k-resolution. A count rate of $10^6$ cps (in the energy range of interest, confined by a dispersive pre-filter) is realistic for future experiments at high-performance fs sources. Under such conditions the intensity array needs an acquisition time of $10^4$ s = 2.8 h. The dichroism asymmetries need pairs of such arrays with orthogonal photon polarizations. Two arrays ($I^R$, $I^L$) acquired with right- and left-circular polarization yield the $A_{CDAD}=(I^R-I^L)/(I^R+I^L)$ and the intensity $I=(I^R+I^L)$ as discussed in detail by Fedchenko et al.[156]. The LDAD and TRDAD asymmetry can be extracted from a single measurement without the need to rearrange the sample geometry or switch the photon-polarization state[242].

In favorable cases like the example in III.A, the count rate in the spin detector can also be of the order of $10^6$ cps. Assuming an average count rate of $10^6$ cps in the spin detector, a 'complete' data set can be captured in the full 3D BZ and 4-6 eV energy interval in about one day. It consists of intensity I, $A_{CDAD}$, $A_{LDAD}$, ($A_{TRDAD}$ if applicable) and the spin-polarization, all recorded within a delay-time interval of several ps (time resolution defined by the source).

A full mapping of the spin-polarization vector **P** is still a big challenge due to the figure-of-merit of spin filters and space-charge effects, which set a practical limit to the effective count rates in the spin filter. A few laboratories (among them again the HEXTOF at FLASH) are starting to work with imaging spin filters in ToF-MMs at fs sources, but there is no systematics yet. The vector **P** can be composed by its Cartesian components $P_{x,y,z}$ or alternatively parametrized by the underlying mechanisms as $P_M$ (in ferromagnets), $P_{Fano}$, $P_{Mott}$, $P_{Rashba}$ or $P_{long}$ (longitudinal spin component appearing for Rashba systems with high Z), depending on the properties of the electronic system. Vectorial measurements can be implemented exploiting symmetry properties of the spinfilter material[105] or by spin rotators[243]. More details on the structure of **P** and the relation between atomic and solid-state photoemission are discussed in the review by Heinzmann and Dil[244]. Recently first spin-resolved measurements with fs time resolution were successful[245, 246].

*Time-resolved ToF-MM* has already reached good visibility in the young and dynamic field of fs-photoemission. Ground-breaking achievements dealt with various aspects in the dynamics of phase transitions (III.F), ultrafast mapping of molecular orbitals, tracking of the intrinsic excitonic wavefunctions, and the electromagnetic dressing of bands[247]. Important future aspects will be the exploitation of various appearances of dichroism (TRDAD, iLDAD)[61,62,234] in time-resolved experiments. The higher photon energy at FELs allows *time-resolved XPD* accessing lattice dynamics, e.g. by selective stimulation of phonon modes. First experiments were successful using the HEXTOF[231,232,84], opening the route to quasi-simultaneous probing of electron and lattice dynamics with fs time resolution.

The restrictions by the space charge of photo-emitted electrons can be pushed to a higher limit using special electron optics (retarding front lens) especially for the soft- and hard X-ray regime[111,112]. In contrast, if the limitation is the maximum possible count rate of the detector for those cases where the incident photon flux sets no limit, it will be equivalent whether the huge data array is filled sequentially or by parallel recording of a momentum and energy interval. Detector schemes with high degree of parallelization based on the delay-line principle[94,95] or pixelated solid-state detectors[96,97] will push the count-rate limit, so that the detector will not be the bottleneck of the total recording efficiency.

Undesired background signals caused by higher-order contributions from undulators and monochromators or the intense low-energy background of secondary and pump-induced electrons can be suppressed by bandpass pre-filters. A *k*-imaging hemispherical analyzer can select any desired bandpass for the electrons entering the ToF section[32]. Alternatively, it can even provide a resolution down to few meV, when operated as single-hemisphere MM without ToF recording[35]. As promising alternative for a much simpler type of bandpass pre-filter, we are testing aberration-reducing multipole deflectors like the asymmetric dodecapole[248].

**Acknowledgements**

One of the authors (G.S.) had the privilege to spend a postdoctoral period in Chuck Fadley's laboratory at the University of Hawaii. Chuck's deep understanding of photoemission and the stimulating atmosphere in the group had a long-lasting influence far beyond the postdoc time. Experiencing one of the early photoelectron spectrometers with multichannel detection at work laid the cornerstone for pursuing maximum parallelization of data recording until today.

The results shown in this paper have been obtained in cooperation of the group at Mainz University with several partners. Sincere thanks go to Katerina Medjanik, Olena Fedchenko, Dmitri Vasilyev, Sergey Babenkov, Steinn Agustsson and Sergey Chernov (Mainz University). We thank Jure Demsar (Mainz University), Kai Rossnagel (Kiel University), Ralph Claessen (Würzburg University), Thomas Allison and Alice Kunin (Stony Brook University) for fruitful discussions on various aspects of this work. Special thanks go to Andreas Oelsner and his team (Surface Concept GmbH, Mainz) and to numerous coworkers in the beamline teams at 10m NIM (BESSY II, Berlin), P04 and P22 (PETRA III, Hamburg) and the HEXTOF team at FLASH (DESY, Hamburg) for excellent cooperation. The ToF-MM and spin-filter development was funded by BMBF (05K16UM1, 05K16UMC, 05K19UM1 and 05K19UM2) and Deutsche Forschungsgemeinschaft DFG (German Research Foundation) through TRR 173–268565370 Spin+X (projects A02, A05), project Scho341/16-1 and TRR 288–422213477 (project B04).

**Data availability statement**

All data shown within this article have been published previously as stated in the figure captions.

**References**

[1] B. Lv, T. Qian, and H. Ding, Angle-resolved photoemission spectroscopy and its application to topological materials, *Nature Rev. Phys*. **1**, 609-626 (2019); doi: 10.1038/s42254-019-0088-5.

[2] I. Belopolski, K. Manna, D. S. Sanchez, G. Chang, B. Ernst, J. Yin, S. S. Zhang, T. Cochran, N. Shumiya, H. Zheng, B. Singh, G. Bian, D. Multer, M. Litskevich, X. Zhou, S.-M. Huang, B. Wang, T.-R. Chang, S.-Y. Xu, A. Bansil, C. Felser, H. Lin, M. Z. Hasan, Discovery of topological Weyl fermion lines and drumhead surface states in a room temperature magnet, *Science* **365**, 1278 (2019).

[3] P. D. C. King, S. Picozzi, R. G. Egdell, and G. Panaccione, Angle, Spin, and Depth Resolved Photoelectron Spectroscopy on Quantum Materials, *Chem. Rev*. **121**, 2816-2856 (2021); doi: 10.1021/acs.chemrev.0c00616.

[4] S. Subramanian, Q. T. Campbell, S. K. Moser, J. Kiemle, P. Zimmermann, P. Seifert, F. Sigger, D. Sharma, H. Al-Sadeg, M. Labella, D. Waters, R. M. Feenstra, R. J. Koch, C. Jozwiak, A. Bostwick, E. Rotenberg, I. Dabo, A. W. Holleitner, T. E. Beechem, U. Wurstbauer, and J. A. Robinson, Photophysics and Electronic Structure of Lateral Graphene/$MoS_2$ and Metal/$MoS_2$ Junctions, *ACS Nano* **14**, 16663–16671 (2020).

[5] P. Hofmann, Accessing the spectral function of in operando devices by angle-resolved photoemission spectroscopy, *AVS Quantum Sci*. **3**, 021101 (2021); doi: 10.1116/5.0038637.

[6] X. Yan, D. D. Fong, H. Zhou, and J. L. McChesney, Synchrotron studies of functional interfaces and the state of the art: A perspective, *J. Appl. Phys.* **129**, 220902 (2021); doi: 10.1063/5.0053291.

[7] S. Lisi, X. Lu, T. Benschop, T. A. de Jong, P. Stepanov, J. R. Duran, F. Margot, I. Cucchi, E. Cappelli, A. Hunter, A. Tamai, V. Kandyba, A. Giampietri, A. Barinov, J. Jobst, V. Stalman, M. Leeuwenhoek, K. Watanabe, T. Taniguchi, L. Rademaker, S. J. van der Molen, M. P. Allan, D. K. Efetov, and F. Baumberger, Observation of flat bands in twisted bilayer graphene, *Nature Physics* **17**, 189 (2021).

[8] K. Siegbahn, U. Gelius, H. Siegbahn, and E. Olson, Angular Distribution of Electrons in ESCA Spectra from a Single Crystal, *Phys. Scripta* **1,** 272 (1970). doi.org/10.1088/0031-8949/1/5-6/017.

[9] C. S. Fadley and S.A.L. Bergstrom, Angular Distributions of Photoelectrons from a Metal Single Crystal, *Phys. Letters* **35A,** 375 (1971); doi.org/10.1016/0375-9601(71)90745-6.

[10] C. S. Fadley, M.A. Van Hove, Z. Hussain, and A.P Kaduwela, Photoelectron diffraction: new dimensions in space, time, and spin *J. Electron Spectrosc. Relat. Phenom.* **75,** 273 (1995); doi.org/10.1016/0368-2048(95)02545-6.

[11] J. Osterwalder, ed. By K. Wandelt, Photoelectron Spectroscopy and Diffraction, in '*Handbook on Surface and Interface Science*', Vol. **1**, 151-214 (2011) (Wiley-VCH, Weinheim); doi.org/10.1002/9783527680535.ch5.

[12] C. Westphal, The study of the local atomic structure by means of X–ray photoelectron diffraction *Surface Science Reports* **50,** 1-106 (2003); doi.org/10.1016/S0167-5729(03)00022-0.

[13] A. Winkelmann, C.S. Fadley, and F.J. Garcia de Abajo, High-energy photoelectron diffraction: model calculations and future possibilities *New J. of Phys*. **10**, 113002 (2008); doi:10.1088/1367-2630/10/11/113002.

[14] D. P. Woodruff, Surface structural information from photoelectron diffraction *J. Electron Spectrosc. Relat. Phenom.* **186** 178-179 (2010); doi.org/10.1016/j.elspec.2009.06.008.

[15] D. P. Woodruff, and A.M. Bradshaw, Adsorbate structure determination on surfaces using photoelectron diffraction, *Rep. Prog. Phys.* **57**, 1029 (1994); and *J. Electron Spectrosc. Relat. Phenom.* **126,** 55–65 (2002); doi.org/10.1103/PhysRevB.47.13941.

[16] T. Kinoshita, T. Muro, T. Matsushita , H. Osawa, T. Ohkochi, F. Matsui, H. Matsuda, M. Shimomura, M. Taguchi, and H. Daimon, Progress in photoelectron holography at SPring-8, *Jap. J. of Appl. Phys*. **58**, 110503 (2019).

[17] T. Matsushita, T. Muro, F. Matsui, N. Happo, S. Hosokawa,K. Ohoyama, A. Sato-Tomita, Y. C. Sasaki, and K. Hayashi, Principle and Reconstruction Algorithm for Atomic-Resolution Holography, *Journal of the Physical Society of Japan* **87**, 061002 (2018).

[18] A. Chassé and T. Chassé, Theory and Application of Photoelectron Diffraction for Complex Oxide Systems, *Journal of the Physical Society of Japan* **87**, 061006 (2018).

[19] P. Krüger, Photoelectron Diffraction from Valence States of Oriented Molecules, *Journal of the Physical Society of Japan* **87**, 061007 (2018).

[20] B. Sinkovic, B. Hermsmeier, and C.S. Fadley, Observation of Spin-Polarized Photoelectron Diffraction, *Phys. Rev. Lett*. **55,** 1227 (1985); doi.org/10.1103/PhysRevLett.55.1227.

[21] B. Hermsmeier, J. Osterwalder, D.J. Friedmann and C.S. Fadley, Evidence for a High-Temperature Short-Range-Magnetic-Order Transition in MnO(001), *Phys. Rev. Lett*. **62**, 478 (1989); doi.org/10.1103/PhysRevLett.62.478.

[22] O. Fedchenko, A. Winkelmann, S. Chernov, K. Medjanik, S. Babenkov, S. Agustsson, D. Vasilyev, M. Hoesch, H.-J. Elmers, and G. Schönhense, Emitter-Site Specificity of Hard X-Ray Photoelectron Kikuchi-Diffraction, *New J. of Phys*. **22**, 103002 (2020); doi: 10.1088/1367-2630/abb68b.

[23] K. Medjanik, O. Fedchenko, O. Yastrubchak, J. Sadowski, M. Sawicki, L. Gluba, D. Vasilyev, S. Babenkov, S. Chernov, A. Winkelmann, H. J. Elmers, and G. Schönhense, Site-specific Atomic Order and Band Structure Tailoring in the Diluted Magnetic Semiconductor (In,Ga,Mn)As, *Phys. Rev.* B **103**, 075107 (2021); doi: 10.1103/PhysRevB. 103.075107.

[24] T. Kiss, T. Shimojima, K. Ishizaka, A. Chainani, T. Togashi, T. Kanai, X.-Y. Wang, C.-T. Chen, S. Watanabe, and S. Shin, A versatile system for ultrahigh resolution, low temperature, and polarization dependent Laser-angle-resolved photoemission spectroscopy, *Rev. Sci. Instrum*. **79**, 023106 (2008).

[25] T. Shimojima , K. Okazaki, and S. Shin, Low-Temperature and High-Energy-Resolution Laser Photo-emission Spectroscopy, *J. Phys. Soc. Jpn*. **84**, 072001 (2015); doi:10.7566/JPSJ.84.072001.

[26] A. Varykhalov, 1$^2$-ARPES: The ultra-high-resolution photoemission station at the U112-PGM-2a-1$^2$ beamline at BESSY II, *Journal of large-scale research facilities*, 4, A128 (2018); http://dx.doi.org/10.17815/jlsrf-4-99.

[27] V. B. Zabolotnyy et al., Renormalized band structure of $Sr_2RuO_4$: A quasiparticle tight-binding approach, *J. Electron Spectrosc. Relat. Phenom*. **191**, 48 (2013).

[28] A. Tamai, W. Meevasana, P. D. C. King, C. W. Nicholson, A. de la Torre, E. Rozbicki, and F. Baumberger, Spin-orbit splitting of the Shockley surface state on Cu(111), *Phys. Rev.* B **87**, 075113 (2013).

[29] B. Krömker, M. Escher, D. Funnemann, D. Hartung, H. Engelhard, and J. Kirschner, Development of a momentum microscope for time resolved band structure imaging, *Rev. Sci. Instrum.* **79**, 053702 (2008).

[30] C. Tusche, A. Krasyuk, and J. Kirschner, Spin resolved bandstructure imaging with a high resolution momentum microscope, *Ultramicr*. **159,** 520 (2015).

[31] C. Tusche, Y.-J. Chen, C. M. Schneider, and J. Kirschner, "Imaging properties of hemispherical electrostatic energy analyzers for high-resolution momentum microscopy", *Ultramicr*. **206,** 112815 (2019)

[32] G. Schönhense, S. Babenkov, D. Vasilyev, H.-J. Elmers, and K. Medjanik, Single-Hemisphere Photoelectron Momentum Microscope with Time-of-Flight Recording, *Rev. Sci. Instrum*. **91**, 123110 (2020); doi: 10.1063/5.0024074.

[33] F. Matsui, S. Makita, H. Matsuda, T. Yano, E. Nakamura, K. Tanaka, S. Suga, and S. Kera, Photoelectron Momentum Microscope at BL6U of UVSOR-III Synchrotron, *Jap. J. of Appl. Phys*. JJAP-102449.R1 (2020).

[34] K. Medjanik, O. Fedchenko, S. Chernov, D. Kutnyakhov, M. Ellguth, A. Oelsner, B. Schönhense, T. R. F. Peixoto, P. Lutz, C.-H. Min, F. Reinert, S. Däster, Y. Acremann, J. Viefhaus, W. Wurth, H. J. Elmers, and G. Schönhense, Direct 3D mapping of the Fermi surface and Fermi velocity, *Nature Mat.* **16**, 615 (2017).

[35] G. Schönhense, K. Medjanik, O. Fedchenko, A. Zymaková, S. Chernov, D. Kutnyakhov, D. Vasilyev, S. Babenkov, H. J. Elmers, P. Baumgärtel, P. Goslawski, G. Öhrwall, T. Grunske, T. Kauerhof, K. von Volkmann, M. Kallmayer, M. Ellguth, and A. Oelsner, Time-of-Flight Photoelectron Momentum Microscopy with 80-500 MHz Photon Sources: Electron-Optical Pulse Picker or Bandpass Pre-Filter, *J. of Synchr. Radiation* **28**, 1891 (2021).

[36] A. Oelsner, O. Schmidt, M. Schicketanz, M.J. Klais, G. Schönhense, V. Mergel, O. Jagutzki, and H. Schmidt-Böcking, Microspectroscopy and imaging using a delayline-detector in time-of-flight photoemission microscopy, *Rev. Sci. Instrum.* **72**, 3968 (2001).

[37] G. Schönhense, A. Oelsner, O. Schmidt, G. H. Fecher, V. Mergel, O. Jagutzki, and H. Schmidt-Böcking, Time-Of-Flight Photoemission Electron Microscopy - A New Way To Chemical Surface Analysis, *Surf. Sci*. **480,** 180-187 (2001).

[38] J. Rossbach, J. R. Schneider, and W. Wurth, 10 years of pioneering X-ray science at the Free-Electron Laser FLASH at DESY, *Physics Reports* **808**, 1–74 (2019); https://doi.org/10.1016/j.physrep. 2019.02.002.

[39] I. Matsuda and Y. Kubota, Recent Progress in Spectroscopies Using Soft X-ray Free-electron Lasers, *Chem. Lett*. **50**, 1336–1344 (2021); doi:10.1246/cl.200881.

[40] https://www.eli-alps.hu/

[41] https://nsf-nexus.osu.edu/about/

[42] D. Kutnyakhov et al., Time- and momentum-resolved photoemission studies using time-of-flight momentum microscopy at a free-electron laser, *Rev. Sci. Instrum*. **91**, 013109 (2020); doi: 10.1063/1.5118777.

[43] M. Dendzik, R. P. Xian, E. Perfetto, D. Sangalli, D. Kutnyakhov, S. Dong, S. Beaulieu, T. Pincelli, F. Pressacco, D. Curcio, S. Y. Agustsson, M. Heber, J. Hauer, W. Wurth, G. Brenner, Y. Acremann, P. Hofmann, M. Wolf, A. Marini, G. Stefanucci, L. Rettig, and R. Ernstorfer, Observation of an Excitonic Mott Transition Through Ultrafast Core-cum-Conduction Photoemission Spectroscopy, *Phys. Rev. Lett*. **125**, 096401 (2020).

[44] F. Pressacco, D. Sangalli, V. Uhlír, D. Kutnyakhov, J. Ander Arregi, S. Y. Agustsson, G. Brenner, H. Redlin, M. Heber, D. Vasilyev, J. Demsar, G. Schönhense, M. Gatti, A. Marini, W. Wurth and F. Sirotti, Subpicosecond metamagnetic phase transition driven by non-equilibrium electron dynamics, *Nature Commun*. **12**, 5088 (2021); doi: 10.1038/s41467-021-25347-3.

[45] https://www.xfel.eu

[46] https://lcls.slac.stanford.edu/lcls-ii

[47] W. S. Fann, R. Storz, H. W. K. Tom, and J. Bokor, *Phys. Rev. Lett*. **68**, 2834 (1992).

[48] W. S. Fann, R. Storz, H. W. K. Tom, and J. Bokor, *Phys. Rev*. B **46**, 13592 (1992).

[49] M. Obergfell and J. Demsar, *Phys. Rev. Lett*. **124**, 037401 (2020).

[50] S. Beaulieu, S. Dong , N. Tancogne-Dejean, M. Dendzik, T. Pincelli , J. Maklar , R. P. Xian , M. A. Sentef, M. Wolf , A. Rubio, L. Rettig, and R. Ernstorfer, Ultrafast dynamical Lifshitz transition, *Science Advances* **7**, eabd9275 (2020).

[51] B. Koopmans, G. Malinowski, F. D. Longa, D. Steiauf, M. Faehnle, T. Roth, M. Cinchetti, and M. Aeschlimann, Explaining the paradoxical diversity of ultrafast laser-induced demagnetization, *Nat. Mater*. **9**, 259 (2010).

[52] M. Sentef, A. F. Kemper, B. Moritz, J. K. Freericks, Z.-X. Shen and T. P. Devereaux, Examining Electron-Boson Coupling Using Time-Resolved Spectroscopy, *Phys. Rev*. X **3**, 041033 (2013).

[53] T. P. Devereaux, A. M. Shvaika, K. Wu, K. Wohlfeld, C. J. Jia, Y. Wang, B. Moritz, L. Chaix, W.-S. Lee, Z.-X. Shen, G. Ghiringhelli, and L. Braicovich, Directly Characterizing the Relative Strength and Momentum Dependence of Electron-Phonon Coupling Using Resonant Inelastic X-Ray Scattering, *Phys. Rev. X* **6**, 041019 (2016).

[54] M. J. Stern, L. P. Rene de Cotret, M. R. Otto, R. P. Chatelain, J.-P. Boisvert, M. Sutton, and B. J. Siwick, Mapping momentum-dependent electron-phonon coupling and nonequilibrium phonon dynamics with ultrafast electron diffuse scattering, *Phys. Rev.* B **97**, 165416 (2018).

[55] J. Braun, J. Minár, and H. Ebert, Correlation, temperature and disorder: Recent developments in the one-step description of angle-resolved photoemission, *Physics Reports* (2018), https://doi.org/10.1016/j.physrep.2018.02.007.

[56] S. Hüfner, *Photoelectron Spectroscopy*, Springer, Berlin-Heidelberg, 2003.

[57] F. Reinert and S. Hüfner, Photoemission spectroscopy—from early days to recent applications, *New J. Phys*. **7**, 97 (2005).

[58] E. W. Plummer and W. Eberhardt, Angle-resolved photoemission as a tool for the study of surfaces, *Adv. In Chem. Phys.* **49**, 533 (1982).

[59] G. Schönhense, Circular Dichroism and Spin Polarization in Photoemission from Adsorbates and Non-Magnetic Solids, *Phys. Scr*. T **T31**, 255 (1990).

[60] N. A. Cherepkov and G. Schönhense, Linear dichroism in photoemission from oriented molecules, *Europhys. Lett.* **24**, 79 (1993).

[61] S. Beaulieu, J. Schusser, S. Dong, M. Schüler, T. Pincelli, M. Dendzik, J. Maklar, A. Neef, H. Ebert, K. Hricovini, M. Wolf, J. Braun, L. Rettig, J. Minár, and R. Ernstorfer, Revealing Hidden Orbital Pseudospin Texture with Time-Reversal Dichroism in Photoelectron Angular Distributions, *Phys. Rev. Lett*. **125**, 216404 (2020).

[62] S. Beaulieu, M. Schüler, J. Schusser, S. Dong, T. Pincelli, J. Maklar, A. Neef, F. Reinert, M. Wolf, L. Rettig, J. Minár, and R. Ernstorfer, Unveiling the Orbital Texture of 1T-TiTe$_2$ using Intrinsic Linear Dichroism in Multidimensional Photoemission Spectroscopy, e-print on *arXiv* 2107.07158 (2021).

[63] U. Fano, Spin Orientation of Photoelectrons Ejected by Circularly Polarized Light, *Phys. Rev*. **178**, 131 (1969).

[64] D. Pierce and F. Meier, Photoemission of spin-polarized electrons from GaAs, *Phys. Rev*. B **13**, 5484 (1976).

[65] U. Heinzmann, G. Schönhense, and J. Kessler, Polarization of photoelectrons ejected by unpolarized light from xenon atoms, *Phys. Rev. Lett*. **42**, 1603 (1979).

[66] G. Schönhense, Angular dependence of the polarization of photoelectrons ejected by plane-polarized radiation from argon and xenon atoms, *Phys. Rev. Lett*. **44**, 640 (1980).

[67] E. Tamura, W. Piepke, and R. Feder, New spin-polarization effect in photoemission from nonmagnetic surfaces, *Phys. Rev. Lett*. **59**, 934 (1987).

[68] N. F. Mott and H. S. W. Massey, *Theory of Atomic Collisions* (Oxford University Press, London, England, 1965).

[69] D. Vasilyev, C. Tusche, F. Giebels, H. Gollisch, R. Feder, and J. Kirschner, Low-energy electron reflection from Au-passivated Ir (001) for application in imaging spin-filters, *J. Electron Spectrosc. Relat. Phenom*. **199**, 10 (2015).

[70] H. Ebert, Fully relativistic band structure calculations for magnetic solids - formalism and application, in: H. Dreyssé (Ed.), *Electronic Structure and Physical Properties of Solids*, 535, Springer, Berlin (2000), p. 191.

[71] J. Braun, The theory of angle-resolved ultraviolet photoemission and its applications to ordered materials, *Rep. Prog. Phys*. **59,** 1267 (1996).

[72] J. Henk, *Handbook of Thin Film Materials* vol 2, ed H S Nalwa (San Diego, CA: Academic) chapter 10, p 479 (2001).

[73] W. Weber, S. Riesen, and D. Oberli, Spin-Dependent Transmission and Spin Precession of Electrons Passing Across Ferromagnets, in *Physics of Low Dimensional Systems*, Ed. by J. L. Morán-López, Kluwer Academic/ Plenum Publishers, New York (2001); https://doi.org/10.1007/0-306-47111-6_33.

[74] J. Kirschner , *Polarized Electrons at Surfaces*, Springer Tracts Mod. Phys., Berlin (1985).

[75] J. Kessler, The 'perfect' photoionization experiment, *Comments Atom. Mol. Phys*. **10,** 47 (1981).

[76] N. A. Cherepkov, Complete experiments in photoionization of atoms and molecules, *Journal of Electron Spectrosc. Relat. Phenom.* **144-147**, 1197–1201 (2005).

[77] G. Schönhense, C. Westphal, J. Bansmann, and M. Getzlaff, Circular Dichroism in Photoemission from Nonmagnetic Low-Z Solids: A Conspicuous Effect of the Photon Spin, *Europhys. Letters* **17**, 727-732 (1992)

[78] G. Schönhense, K. Medjanik, S. Chernov, D. Kutnyakhov, O. Fedchenko, M. Ellguth, D.Vasilyev, A. Zaporozhchenko-Zymaková, D. Panzer, A. Oelsner, C. Tusche, B. Schönhense, J. Braun, J. Minár, H. Ebert, J. Viefhaus, W. Wurth, and H. J. Elmers, Spin-Filtered Time-of-Flight k-Space Microscopy of Ir – Towards the 'Complete' Photoemission Experiment, *Ultramicr*. **183**, 19 (2017).

[79] K. Fehre et al., Fourfold Differential Photoelectron Circular Dichroism, *Phys. Rev. Lett*. **127**, 103201 (2021).

[80] R. Dörner, V. Mergel, O. Jagutzki, L. Spielberger, J. Ullrich, R. Moshammer, and H. Schmidt-Böcking, *Phys. Rep*. **330**, 95 (2000).

[81] M. Scholz, K. Baumgärtner, C. Metzger, D. Kutnyakhov, M. Heber, C. H. Min, T. R. F. Peixoto, M. Reiser, C. Kim, W. Lu, R. Shayduk, W. M. Izquierdo, G. Brenner, F. Roth, F. Pressacco, A. Schöll, S. Molodtsov, W. Wurth, F. Reinert, and A. Madsen, Ultrafast molecular orbital imaging of a pentacene thin film using a free electron laser, e-print *arXiv*: 1907.10434 (2019).

[82] G. S. M. Jansen, M. Keunecke, M. Düvel, C. Möller, D. Schmitt, W. Bennecke, F. J. S. Kappert, D. Steil, D. R. Luke, S. Steil, and S. Mathias, Efficient orbital imaging based on ultrafast momentum microscopy and sparsity-driven phase retrieval, *New J. of Phys*. **22**, 063012 (2020).

[83] R. Wallauer, M. Raths, K. Stallberg, L. Münster, D. Brandstetter, X. Yang, J. Güdde, P. Puschnig, S. Soubatch, C. Kumpf, F. C. Bocquet, F. S. Tautz, and U. Höfer, Tracing orbital images on ultrafast time scales, *Science* **371**, 1056 (2021).

[84] K. Baumgärtner, M. Nozaki, C. Metzger, N. Wind, M. Haniuda, M. Heber, D. Kutnyakhov, F. Pressaco, L. Wenthaus, F. Roth, C.-H. Min, F. Reinert, A. Madsen, S. Mahatha, K. Niki, K. Rossnagel, and M. Scholz, The complete molecular movie: Capturing electronic and structural dynamics at the molecule-2D material interface, *in prep.*

[85] S. Dong, M. Puppin, T. Pincelli, S. Beaulieu, D. Christiansen, H. Hübener, C. W. Nicholson, R. P. Xian, M. Dendzik, Y. Deng, Y. W. Windsor, M. Selig, E. Malic, A. Rubio, A. Knorr, M. Wolf, L. Rettig, and R. Ernstorfer, Direct measurement of key exciton properties: energy, dynamics and spatial distribution of the wave function, *Natural Sci*. **1**, e10010 (2021); doi: 0.1002/ntls.10010.

[86] M. K. L. Man, J. Madéo, C. Sahoo, K. Xie, M. Campbell, V. Pareek, A. Karmakar, E. Laine Wong, A. Al-Mahboob, N. S. Chan, D, R. Bacon, X. Zhu, M. M. M. Abdelrasoul, X. Li, T. F. Heinz, F. H. da Jornada, T. Cao, and K. M. Dani, Experimental measurement of the intrinsic excitonic wave function, *Science Adv.* **7**, eabg0192 (2021).

[87] M. Schüler, T. Pincelli, S. Dong, T. P. Devereaux, M. Wolf, L. Rettig, R. Ernstorfer, and S. Beaulieu, Bloch Wavefunction Reconstruction using Multidimensional Photoemission Spectroscopy, e-print *arXiv* 2103.17168 (2021).

[88] O. Karni, E. Barré, V. Pareek, J. D. Georgaras, M. K. L. Man, C. Sahoo, D. R. Bacon, X. Zhu, H. B. Ribeiro, A. L. O'Beirne, J. Hu, A. Al-Mahboob, M. M. M. Abdelrasoul, N. S. Chan, A. Karmakar, A. J. Winchester, B. Kim, K. Watanabe, T. Taniguchi, K. Barmak, J. Madéo, F. H. da Jornada, T. F. Heinz, and K. M. Dani, Moiré-localized interlayer exciton wavefunctions captured by imaging its electron and hole constituents, e-print *arXiv* 2108.01933 (2021).

[89] D. Schmitt, J. P. Bange, W. Bennecke, A. A. AlMutairi, K. Watanabe, T. Taniguchi, D. Steil, D. R. Luke, R. T. Weitz, S. Steil, G. S. M. Jansen, S. Hofmann, M. Reutzel, and S. Mathias, Formation of Moiré interlayer excitons in space and time, e-print *arXiv* 2112.05011 (2021).

[90] R. Wallauer, R Perea-Causin, L. Münster, S. Zajusch, S. Brem, J. Güdde, K. Tanimura, K.-Q. Lin, R. Huber, E. Malic, and U. Höfer, Momentum-Resolved Observation of Exciton Formation Dynamics in Monolayer $WS_2$, *Nano Letters* (2021); doi.10.1021/acs.nanolett. 1c01839.

[91] A. Oelsner, M. Rohmer, Ch. Schneider, D. Bayer, G. Schönhense, and M. Aeschlimann, Time- and Energy resolved photoemission electron microscopy – imaging of photoelectron time-of-flight analysis by means of pulsed excitations, *J. Electron Spectrosc. Relat. Phenom*. **178-179,** 317-330 (2010).

[92] M. H. Berntsen, O. Götber, and O. Tjernberg, An experimental setup for high resolution 10.5 eV laser-based angle-resolved photoelectron spectroscopy using a time-of-flight electron analyser, *Rev. Sci. Instrum*. **82**, 095113 (2011).

[93] R. Ovsyannikov, P. Karlsson, M. Lundqvist, C. Lupulescu, W. Eberhardt, A. Föhlisch, S. Svensson, and N. Martensson, Principles and operation of a new type of electron spectrometer-ArTOF, *J. Electron Spectrosc. Relat. Phenom*. **191**, 92-103 (2013).

[94] https://www.surface-concept. com/ downloads/info/ml_dld.pdf

[95] https://www.roentdek.com/detectors/

[96] A. Zhao, M. van Beuzekom, B. Bouwens, D. Byelov, I. Chakaberia, C. Cheng, E. Maddox, A. Nomerotski, P. Svihra, J. Visser, V. Vrba, and T. Weinacht, Coincidence velocity map imaging using Tpx3Cam, a time stamping optical camera with 1.5 ns timing resolution, *Rev. Sci. Instrum*. **88**, 113104 (2017); doi: 10.1063/1.4996888.

[97] G. Giacomini, W. Chen, F. Lanni, and A. Tricoli, Development of a technology for the fabrication of Low-Gain Avalanche Diodes at BNL, *Nucl. Instr. Meth*. A **934**, 52 (2019); doi: 10.1016/ j.nima.2019.04.073.

[98] M. Kolbe, P. Lushchyk, B. Petereit, H.J. Elmers, G. Schönhense, A. Oelsner, C. Tusche, and J. Kirschner, Highly Efficient Multichannel Spin-Polarization Detection, *Phys. Rev. Lett*. **107,** 207601 (2011).

[99] C. Tusche, M. Ellguth, A. A. Ünal, C. T. Chiang, A. Winkelmann, A. Krasyuk, M. Hahn, G. Schönhense, and J. Kirschner, Spin resolved photoelectron microscopy using a two-dimensional spin-polarizing electron mirror, *Appl. Phys. Lett*. **99,** 032505 (2011).

[100] S. Suga and C. Tusche, Photoelectron spectroscopy in a wide hν region from 6 eV to 8 keV with full momentum and spin resolution, *J. Electron Spectrosc. Relat. Phenom*. **200,** 119–142 (2015).

[101] D. Kutnyakhov, P. Lushchyk, D. Perriard, M. Kolbe, K. Medjanik, E. Fedchenko, S.A. Nepijko, H.J. Elmers, G. Salvatella, R. Gort, T. Bähler, T. Michlmayer, A. Fognini, Y. Acremann, A. Vaterlaus, C. Tusche, A. Krasyuk, J. Kirschner, F. Giebels, H. Gollisch, R. Feder, and G. Schönhense, Imaging Spin Filter for Electrons Based on Specular Reflection from Ir (001), *Ultramicrosc*. **130,** 63–69 (2013).

[102] J. Kirschner, F. Giebels, H. Gollisch, and R. Feder, Spin-polarized electron scattering from pseudomorphic Au on Ir(001), *Phys. Rev*. B **88**, 125419 (2013).

[103] G. K. Gotlieb, Z. Hussain, A. Bostwick, A. Lanzara, and C. Jozwiak, Rapid high-resolution spin- and angle-resolved photoemission spectroscopy with pulsed laser source and time-of-flight spectrometer, *Rev. Sci. Instrum*. **84**, 093904 (2013).

[104] D. Kutnyakhov, H.J. Elmers, G. Schönhense, C. Tusche, S. Borek, J. Braun, J. Minàr, and H. Ebert, Specular reflection of spin-polarized electrons from a W(001) spin-filter crystal in a larger range of scattering energies and angles, *Phys. Rev*. B **91,** 014416 (2015).

[105] E. D. Schäfer, S. Borek, J. Braun, J. Minar, H. Ebert, K. Medjanik, D. Kutnyakhov, G. Schönhense, and H.-J. Elmers, Vectorial spin-polarization detection in multichannel spin-resolved photoemission spectroscopy using an Ir(001) imaging spin filter, *Phys. Rev*. B **95**, 104423 (2017).

[106] F. Ji, T. Shi, M. Ye, W. Wan, Z. Liu, J. Wang, T. Xu, and S. Qiao, Multichannel Exchange-Scattering Spin Polarimetry, *Phys. Rev. Lett*. **116**, 177601 (2016).

[107] C. Angrick, J. Braun, H. Ebert, and M. Donath, Spin-dependent electron reflection at W(110), *J. Phys.: Condens. Matter* **33**, 115001 (2020).

[108] J. Kessler, *Polarized Electrons,* Springer, Berlin (2nd ed. 1985); ISBN 978-3-662-02434-8.

[109] S. Chernov, C. Lidig, O. Fedchenko, K. Medjanik, S. Babenkov, D. Vasilyev, M. Jourdan, G. Schönhense, and H. J. Elmers, Band structure tuning of Heusler compounds: Spin- and momentum-resolved electronic structure analysis of compounds with different band filling, *Phys. Rev*. B **103**, 054407 (2021).

[110] C. Tusche, M. Ellguth, A. Krasyuk, A. Winkelmann, D. Kutnyakhov, P. Lushchyk, K. Medjanik, G. Schönhense, and J. Kirschner, Quantitative spin polarization analysis in photoelectron emission microscopy with an imaging spin filter, *Ultramic*. **130**, 70 (2013).

[111] B. Schönhense, K. Medjanik, O. Fedchenko, S. Chernov, M. Ellguth, D. Vasilyev, A. Oelsner, J. Viefhaus, D. Kutnyakhov, W. Wurth, H. J. Elmers, and G. Schönhense, Multidimensional Photoemission Spectroscopy - the Space-Charge Limit, *New J. of Physics* **20**, 033004 (2018).

[112] G. Schönhense, D. Kutnyakhov, F. Pressacco, M. Heber, N. Wind, S. Y. Agustsson, S. Babenkov, D. Vasilyev, O. Fedchenko, S. Chernov, L. Rettig, B. Schönhense, L. Wenthaus, G. Brenner, S. Dziarzhytski, S. Palutke, S. K. Mahatha, N. Schirmel, H. Redlin, B. Manschwetus, I. Hartl, Yu. Matveyev, A. Gloskovskii, C. Schlueter, V. Shokeen, H. Duerr, T. K. Allison, M. Beye, K. Rossnagel, H. J. Elmers, and K. Medjanik, *S*uppression of the vacuum space-charge effect in fs-photoemission by a retarding electrostatic front lens, *Rev. Sci. Instrum*. **92** (2021); doi: 10.1063/5.0046567.

[113] A. K. Mills, T. J. Hammond, M. H. C. Lam, and D. J. Jones, *Journal of Physics B: Atomic, Molecular and Optical Physics* **45**, 142001 (2012).

[114] H. Carstens, M. Högner, T. Saule, S. Holzberger, N. Lilienfein, A. Guggenmos, C. Jocher, T. Eidam, D. Esser, V. Tosa, V. Pervak, J. Limpert, A. Tünnermann, U. Kleineberg, F. Krausz, and I. Pupeza, *Optica* **3**, 366 (2016).

[115] O. Kfir, P. Grychtol, E. Turgut, R. Knut, D. Zusin, D. Popmintchev, T. Popmintchev, H. Nembach, J. M. Shaw, A. Fleischer, H. Kapteyn, M. Murnane, and O. Cohen, Generation of bright phase-matched circularly-polarized extreme ultraviolet high harmonics, *Nature Photonics* **9**, 99–105 (2015).

[116] H. Höchst, R. Patel, and F. Middleton, Multiple-reflection λ/4 phase shifter: A viable alternative to generate circular-polarized synchrotron radiation, *Nucl. Instr. Meth. in Phys. Research* A **347**, 107-114 (1994).

[117] M. P. Seah and W. A. Dench, Quantitative electron spectroscopy of surfaces: A standard data base for electron inelastic mean free paths in solids, *Surf. Interface Anal*. **1**, 2 (1979).

[118] C. Lidig, J. Minár, J. Braun, H. Ebert, A. Gloskovskii, J. A. Krieger, V. Strocov, M. Kläui, and M. Jourdan, Surface resonance of thin films of the Heusler half-metal $Co_2MnSi$ probed by soft x-ray angular resolved photoemission spectroscopy, *Phys. Rev*. B **99**, 174432 (2019).

[119] S. Suga and A. Sekiyama, *Photoelectron Spectroscopy. Bulk and Surface Electronic Structures* (Springer, Berlin, 2014).

[120] J. Braun, M. Jourdan, A. Kronenberg, S. Chadov, B. Balke, M. Kolbe, A. Gloskovskii, H. J. Elmers, G. Schönhense, C. Felser, M. Kläui, H. Ebert, and J. Minár, Monitoring surface resonances on $Co_2MnSi$(100) by spin-resolved photoelectron spectroscopy, *Phys. Rev.* B **91**, 195128 (2015).

[121] J. Ma, V. I. Hegde, K. Munira, Y. Xie, S. Keshavarz, D. T. Mildebrath, C. Wolverton, A. W. Ghosh, and W. H. Butler, Computational investigation of half-Heusler compounds for spintronics applications, *Phys. Rev*. B **95**, 024411 (2017); http://heusleralloys.mint.ua.edu/

[122] M. Hoesch, M. Muntwiler, V. N. Petrov, M. Hengsberger, L. Patthey, M. Shi, M. Falub, T. Greber, and J. Osterwalder, *Phys. Rev*. B **69**, 241401(R) (2004).

[123] F. Reinert, G. Nicolay, S. Schmidt, D. Ehm, and S. Hüfner, *Phys. Rev*. B **63**, 115415 (2001).

[124] S. LaShell, B. A. McDougall, and E. Jensen, *Phys. Rev. Lett*. **77**, 3419 (1996).

[125] E. I. Rashba, *Sov. Phys. Solid State* **2**, 1109 (1960).

[126] F. Reinert, *J. Phys. Cond. Mat*. **15**, S693 (2003).

[127] B. Yan, B. Stadtmueller, N. Haag, S. Jakobs, J. Seidel, D. Jungkenn, S. Mathias, M. Cinchetti, M. Aeschlimann, and C. Felser, *Nat. Commun*. **6**, 10167 (2015).

[128] K. Miyamoto, A. Kimura, K. Kuroda, T. Okuda, K. Shimada, H. Namatame, M. Taniguchi, and M. Donath, *Phys. Rev. Lett*. **108**, 066808 (2012).

[129] A. Varykhalov, D. Marchenko, J. Sanchez-Barriga, E. Golias, O. Rader, and G. Bihlmayer, *Phys. Rev.* B **95**, 245421 (2017).

[130] D. Kutnyakhov, S. Chernov, K. Medjanik, R. Wallauer, C. Tusche, M. Ellguth, S. A. Nepijko, M. Krivenkov, J. Braun, S. Borek, J. Minar, H. Ebert, H. J. Elmers, and G. Schönhense, *Sci. Rep*. **6**, 29394 (2016).

[131] K. Honma, T. Sato, S. Souma, K. Sugawara, Y. Tanaka, and T. Takahashi, *Phys. Rev. Lett*. **115**, 266401 (2015).

[132] H. Mirhosseini, F. Giebels, H. Gollisch, J. Henk, and R. Feder, *New J. Phys*. **15**, 095017 (2013).

[133] A. G. Rybkin, E. E. Krasovskii, D. Marchenko, E. V. Chulkov, A. Varykhalov, O. Rader, and A. M. Shikin, *Phys. Rev.* B **86**, 035117 (2012).

[134] J. Braun, K. Miyamoto, A. Kimura, T. Okuda, M. Donath, H. Ebert, and J. Minar, New J. Phys. **16**, 015005 (2014).

[135] H. J. Elmers, J. Regel, T. Mashof, J. Braun, S. Babenkov, S. Chernov, O. Fedchenko, K. Medjanik, D. Vasilyev, J. Minar, H. Ebert, and G. Schönhense, Rashba splitting of the Tamm surface state on Re(0001) observed by spin-resolved photoemission and scanning tunnelling spectroscopy, *Phys. Rev. Research* **2**, 013296 (2020).

[136] A. Urru and A. Dal Corso, *Surf. Sci*. **686**, 22 (2019).

[137] J. Regel, T. Mashoff, and H. J. Elmers, *Phys. Rev*. B **102**, 115404 (2020).

[138] H. J. Elmers et al., Spin mapping of surface and bulk Rashba states in ferroelectric α-GeTe (111) films, *Phys. Rev*. B **94**, 201403(R) (2016).

[139] S. Tanuma, C. J. Powell, and D.R. Penn, Calculations of electron inelastic mean free paths. IX. Data for 41 elemental solids over the 50 eV to 30 keV range, *Surf. Interface Anal*. **43**, 689 (2011).

[140] J. Viefhaus, F. Scholz, S. Deinert, L. Glaser, M. Ilchen, J. Seltmann, P. Walter, and F. Siewert, The Variable Polarization XUV Beamline P04 at PETRAIII: Optics, Mechanics and their Performance, *Nucl. Instrum. Meth*., **710**, 151 (2013).

[141] V. N. Strocov, X.Wang, M. Shi, M. Kobayashi, J. Krempasky, C. Hess, T. Schmitt, and L. Patthey, Soft-X-Ray ARPES Facility at the ADRESS Beamline of the SLS: Concepts, Technical Realisation and Scientific Applications, *J. Synchrotron Radiat*. **21**, 32 (2014).

[142] F. Polack, M. Silly, C. Chauvet, B. Lagarde, N. Bergeard, M. Izquierdo, O. Chubar, D. Krizmancic, M. Ribbens, J.-P. Duval, C. Basset, S. Kubsky, and F. Sirotti, TEMPO: a New Insertion Device Beamline at SOLEIL for Time Resolved Photoelectron Spectroscopy Experiments on Solids and Interfaces, *AIP Conference Proceedings* **1234**, 185 (2010); doi:10.1063/1.3463169

[143] F. Z. Guo et al., Characterization of spectroscopic photoemission and low energy electron microscope using multipolarized soft x rays at BL17SU/SPring-8, *Rev. Sci. Instrum*. **78**, 066107 (2007); doi: 10.1063/1.2748387.

[144] Y. Saitoh et al., Performance upgrade in the JAEA actinide science beamline BL23SU at SPring-8 with a new twin-helical undulator, *J. Synchrotron Rad.* **19**, 388–393 (2012); doi:10.1107/S0909049512006772Y.

[145] Y. Senba et al., Upgrade of beamline BL25SU for soft x-ray imaging and spectroscopy of solid using nano- and micro-focused beams at SPring-8, *AIP Conference Proceedings* **1741**, 030044 (2016); doi10.1063/1.4952867.

[146] T.-L. Lee and D. A. Duncan, A Two-Color Beamline for Electron Spectroscopies at Diamond Light Source, *Synchr. Rad. News* **31**, 16 (2018).

[147] P. Sjoblom, G. Todorescua, and S. Urpelainen, Understanding the mechanical limitations of the performance of soft X-ray monochromators at MAX IV laboratory, *J. Synchr. Radiation* **27,** 272-283 (2020).

[148] M. B. Trzhaskovskaya and V. G.Yarzhemsky, Dirac-Fock photoionization parameters for HAXPES applications. *Nucl. Data Tables* **119**, 99–174 (2018).

[149] K. Medjanik, S. V. Babenkov, S. Chernov, D. Vasilyev, B. Schönhense, C. Schlueter, A. Gloskovskii, Y. Matveyev, W. Drube, H. J. Elmers, and G. Schönhense, Progress in HAXPES performance combining full-field k-imaging with time-of-flight recording, *J. Synchrotron Rad*. (2019). 26, 1996-2012

[150] A. Gray, Chapter *Hard X-ray Angle-Resolved Photoelectron Spectroscopy (HARPES), pp. 141-157* in 'Hard X-ray Photoelectron Spectroscopy (HAXPES)', ed. by. J. Woicik, Springer (2016); ISBN 978-3-319-24043-5, doi: 10.1007/978-3-319-24043-5_8

[151] J. B. Pendry, Theory of photoemission. *Surf. Sci*. **57**, 679–705 (1976).

[152] J. B. Pendry and J. F. L. Hopkinson, Theory of photoemission. J*. Phys. Colloques* **39**, C4-142–C4-148 (1978).

[153] J. F. L. Hopkinson, J. B. Pendry, and D. J. Titterington, Calculation of photoemission spectra for surfaces of solids. *Comput. Phys. Commun*. **19**, 69–92 (1980).

[154] J. B. Pendry, *Low Energy Electron Diffraction,* Academic Press, London (1974).

[155] G. Schönhense, K. Medjanik, S. Babenkov, D. Vasilyev, M. Ellguth, O. Fedchenko, S. Chernov, B. Schönhense, and H.-J. Elmers, Momentum-Transfer Model of Valence-Band Photoelectron Diffraction, *Comms. Phys*. **3**, 45 (2020).

[156] O. Fedchenko, K. Medjanik, S. Chernov, D. Kutnyakhov, M. Ellguth, A. Oelsner, B. Schönhense, T. Peixoto, P. Lutz, C.-H. Min, F. Reinert, S. Däster, Y. Acremann, J. Viefhaus, W. Wurth, J. Braun, J. Minár, H. Ebert, H. J. Elmers, and G. Schönhense, 4D texture of circular dichroism in soft-x-ray photoemission from tungsten, *New J. of Phys*. **21**, 013017 (2019).

[157] M. B. Nielsen, Z. Li, S. Lizzit, A. Goldoni, and P. Hofmann, Bulk Fermi surface mapping with high-energy angle-resolved photoemission, *J. Phys.: Condens. Matter* **15**, 6919 (2003);

[158] J. N. Nelson, J. P. Ruf, Y. Lee, C. Zeledon, J. K. Kawasaki, S. Moser, C. Jozwiak, E. Rotenberg, A. Bostwick, D. G. Schlom, K. M. Shen, and L. Moreschini, Dirac nodal lines protected against spin-orbit interaction in $IrO_2$, *Phys. Rev. Mat.* **3**, 064205 (2019); doi: 10.1103/PhysRevMaterials.3.064205.

[159] D. Vasilyev, K. Medjanik, S. Babenkov, M. Ellguth, G. Schönhense, and H.-J. Elmers, Relation between spin–orbit induced spin polarization, Fano-effect and circular dichroism in soft x-ray photoemission, *J. Phys.: Condens. Matter* **32**, 135501 (2019).

[160] C. Schlueter, A. Gloskovskii, K. Ederer, S. Piec, M. Sing, R. Claessen, C. Wiemann, C.M. Schneider, K. Medjanik, G. Schönhense, P. Amann, A. Nilsson, and W. Drube, New HAXPES Applications at PETRA III, *AIP Conferences Proceedings* **2054**, 040010 (2019).

[161] A. X. Gray, J. Minár, S. Ueda, P. R. Stone, Y. Yamashita, J. Fujii, J. Braun, L. Plucinski, C. M. Schneider, G. Panaccione, H. Ebert, O. D. Dubon, K. Kobayashi, and C. S. Fadley, Bulk electronic structure of the dilute magnetic semiconductor $Ga_{1-x}Mn_xAs$ through hard X-ray angle-resolved photoemission, *Nat. Mater*. **11**, 957 (2012).

[162] A. X. Gray, C. Papp, S. Ueda, B. Balke, Y. Yamashita, L. Plucinski, J. Minár, J. Braun, E. R. Ylvisaker, C. M. Schneider, W. E. Pickett, H. Ebert, K. Kobayashi, and C. S. Fadley, Probing bulk electronic structure with hard X-ray angle-resolved photoemission, *Nat. Mater*. **10**, 759 (2011).

[163] S. Y. Agustsson, S. V. Chernov, K. Medjanik, S. Babenkov, O. Fedchenko, D. Vasilyev, C. Schlueter, A. Hloskovskii, Yu. Matveyev, K. Kliemt, C. Krellner, J. Demsar, G. Schönhense, and H.-J. Elmers, Temperature-dependent Change of the Electronic Structure in the Kondo Lattice System YbRh$_2$Si$_2$, *J. Phys. Condens. Matter* **33**, 205601 (2021).

[164] Y. Kayanuma, Chapter *Recoil Effects in X-ray Photoelectron Spectroscopy*, pp. 175-195 in 'Hard X-ray Photoelectron Spectroscopy (HAXPES)', ed. by. J. Woicik, Springer, 2016; ISBN 978-3-319-24043-5, doi: 10.1007/978-3-319-24043-5_8

[165] B. Ghomashi, N. Douguet, and L. Argenti, Attosecond Intramolecular Scattering and Vibronic Delays, *Phys. Rev. Lett.* **127**, 203201 (2021).

[166] E. Kukk, D. Ceolin, O. Travnikova, R. Püttner, M. N. Piancastelli, R. Guillemin, L. Journel, T. Marchenko, I. Ismail, J. Martins, J.-P. Rueff, and M. Simon, Unified treatment of recoil and Doppler broadening in molecular high-energy photoemission, *New J. Phys.* **23**, 063077 (2021).

[167] S. Kikuchi, *Proceedings of the Imperial Academy*. **4**, 354–356 (1928).

[168] D. B. Williams and C. B. Carter, *Transmission Electron Microscopy*, Springer Boston, (2009), Chap. 11 Diffraction in TEM, p. 197 ff.

[169] S. Babenkov, K. Medjanik, D. Vasilyev, S. Chernov, C. Schlueter, A. Gloskovskii, Yu. Matveyev, W. Drube, B. Schönhense, K. Rossnagel, H.-J. Elmers, and G. Schönhense, High-accuracy bulk electronic bandmapping with eliminated diffraction effects using hard X-ray photoelectron momentum microscopy, *Comms. Phys*. **2**, 107 (2019); doi: 10.1038/s42005-019-0208-7.

[170] A. Winkelmann, C. S. Fadley, and F. J. Garcia de Abajo, *New J. Phys*. **10**, 113002 (2008); doi:10.1088/1367-2630/10/11/113002.

[171] A. Winkelmann and M. Vos, Site-Specific Recoil Diffraction of Backscattered Electrons in Crystals, *Phys. Rev. Lett.* **106**, 085503 (2011); doi:10.1103/PhysRevLett.106.085503

[172] B. Kaufman and H. J. Lipkin, Momentum transfer to atoms bound in a crystal, *Ann. of Phys*. **18**, 294 (1962); doi: 10.1016/0003-4916(62)90072-6.

[173] Z. I. Wang, Thermal diffuse scattering in sub-angstrom quantitative electron microscopy—phenomenon, effects and approaches, *Micron* **34,** 141-155 (2003).

[174] M. von Laue, *Materiewellen und ihre Interferenzen*, Akademische Verlagsgesellschaft Geest & Portig, Leipzig, (1948).

[175] F. Steglich, J. Aarts, C.D. Bredl, W. Lieke, D. Meschede, W. Franz, and H. Schäfer, Superconductivity in the Presence of Strong Pauli Paramagnetism: CeCu$_2$Si$_2$, *Phys. Rev. Lett*. **43**, 1892 (1979).

[176] S. Nakamura, K. Hyodo, Y. Matsumoto, Y. Haga, H. Sato, S. Ueda, K. Mimura, K. Saiki, K. Iso, M. Yamashita, S. Kittaka, T. Sakakibara, and S. Ohara, *J. Phys. Soc. Japan* **89,** 024705 (2020).

[177] S. Danzenbächer, D. V. Vyalikh, K. Kummer, C. Krellner, M. Holder, M. Höppner, Yu. Kucherenko, C. Geibel, M. Shi, L. Patthey, S. L. Molodtsov, and C. Laubschat, *Phys. Rev. Lett*. **107,** 267601 (2011).

[178] M. Güttler, A. Generalov, S. I. Fujimori, K. Kummer, A. Chikina, S. Seiro, S. Danzenbächer, Yu. M. Koroteev, E. V. Chulkov, M. Radovic, M. Shi, N. C. Plumb, C. Laubschat, J. W. Allen, C. Krellner, C. Geibel, and D. Vyalikh. *Nat. Commun*. **10,** 796 (2019).

[179] P. W. Anderson, *Phys. Rev*. **124**, 41 (1961).

[180] H.C. Choi, B.I. Min, J.H. Shim, K. Haule, and G. Kotliar, *Phys. Rev. Lett*. **108**, 016402 (2012).

[181] D. V. Vyalikh, S. Danzenbächer, Yu. Kucherenko, K. Kummer, C. Krellner, C. Geibel, M. G. Holder, T. K. Kim, C. Laubschat, M. Shi, L. Patthey, R. Follath, and S. L. Molodtsov, *Phys. Rev. Lett*. **105**, 237601 (2010).

[182] D. V. Vyalikh, S. Danzenbächer, A. N. Yaresko, M. Holder, Yu. Kucherenko, C. Laubschat, C. Krellner, Z. Hossain, C. Geibel, M. Shi, L. Patthey, and S. L. Molodtsov, *Phys. Rev. Lett*. **100**, 056402 (2008).

[183] S. Danzenbächer, Yu. Kucherenko, D. V. Vyalikh, M. Holder, C. Laubschat, A. N. Yaresko, C. Krellner, Z. Hossain, C. Geibel, X. J. Zhou, W. L. Yang, N. Mannella, Z. Hussain, Z.-X. Shen, M. Shi, L. Patthey, and S. L. Molodtsov, *Phys. Rev*. B **75**, 045109 (2007).

[184] K. Kummer, S. Patil, A. Chikina, M. Güttler, M. Höppner, A. Generalov, S. Danzenbächer, S. Seiro, A. Hannaske, C. Krellner, Yu. Kucherenko, M. Shi, M. Radovic, E. Rienks, G. Zwicknagl, K. Matho, J. W. Allen, C. Laubschat, C. Geibel, and D. V. Vyalikh, *Phys. Rev*. X **5**, 011028 (2015).

[185] O. Trovarelli, C. Geibel, S. Mederle, C. Langhammer, F. M. Grosche, P. Gegenwart, M. Lang, G. Sparn, and F. Steglich, *Phys. Rev. Lett*. **85,** 626 (2000).

[186] S. Ernst, S. Kirchner, C. Krellner, C. Geibel, G. Zwicknagl, F. Steglich, and S. Wirth, *Nature* **474**, 362 (2011).

[187] A. J. Wilkinson and T.B. Britton, *Mater. Today* **15**, 366 (2012).

[188] M. Ojima, Y. Adachi, S. Suzuki, and T. Yo, *Acta Mater*. **V59**, 4177 (2011).

[189] H.-H. Kim, S. M. Souliou, M. E. Barber, E. Lefrançois, M. Minola, M. Tortora, R. Heid, N. Nandi, R. A. Borzi, G. Garbarino, A. Bosak, J. Porras, T. Loew, M. König, P. J. W. Moll, A. P. Mackenzie, B. Keimer, C. W. Hicks, and M. LeTacon, *Science* **362**, 1040 (2018).

[190] G. Song, V. Borisov, W. R. Meier, M. Xu, K. J. Dusoe, J, T. Sypek, R. Valentí, P. C. Canfield, and S.-W. Lee, *APL Mater*. **7**, 061104 (2019).

[191] S. Kimura, J. Sichelschmidt, J. Ferstl, C. Krellner, C. Geibel, and F. Steglich, *Phys. Rev*. B **74**, 132408 (2006).

[192] D. Leuenberger, J. A. Sobota, S. L. Yang, H. Pfau, D. J. Kim, S.K. Mo, Z. Fisk, P. S. Kirchmann, and Z. X. Shen, *Phys. Rev.* B **97**, 165108 (2018).

[193] H. J. Elmers, S. V. Chernov, S. W. D'Souza, S. P. Bommanaboyena, S. Y. Bodnar, K. Medjanik, S. Babenkov, O. Fedchenko, D. Vasilyev, S. Y. Agustsson, C. Schlueter, A. Gloskovskii, Y. Matveyev, V. N. Strocov, Y. Skourski, L. Smejkal, J. Sinova, J. Minar, M. Kläui, G. Schönhense, and M. Jourdan, *ACS Nano* **14**, 17554 (2020).

[194] Q. Wu and M. Altman, *Ultramicroscopy* **159**, 530 (2015).

[195] T. Pincelli, V. N. Petrov, G. Brajnik, R. Ciprian, V. Lollobrigida, P. Torelli, D. Krizmancic, F. Salvador, A. De Luisa, R. Sergo, A. Gubertini, G. Cautero, S. Carrato, G. Rossi, and G. Panaccione, *Rev. Sci. Instrum*. **87**, 035111 (2016).

[196] J. Krempaský, S. Muff, J. Minár, N. Pilet, M. Fanciulli, A. P. Weber, E. B. Guedes, M. Caputo, E. Müller, V. V. Volobuev, M. Gmitra, C. A. F. Vaz, V. Scagnoli, G. Springholz, and J. H. Dil, Operando imaging of all-electric spin texture manipulation in ferroelectric and multiferroic Rashba semiconductors, *Phys. Rev*. X **8**, 021067 (2018)

[197] S. Ueda and Y. Sakuraba, Direct observation of spin-resolved valence band electronic states from a buried magnetic layer with hard X-ray photoemission, *Science and Technology of Advanced Materials* **22**, 317 (2021); https://doi.org/10.1080/14686996.2021.1912576.

[198] X. Kozina, E. Ikenaga, C.E.V. Barbosa, S. Ouardi, J. Karel, M. Yamamoto, K. Kobayashi, H.J. Elmers, G. Schönhense, and C. Felser, *Electron Spectrosc. Relat. Phenom*. **211**, 12 (2016).

[199] M. Schmitt, O. Kirilmaz, S. Chernov, S. Babenkov, D. Vasilyev, O. Fedchenko, K. Medjanik, Yu. Matveyev, A. Gloskovskii, C. Schlueter, A. Winkelmann, L. Dudy, H.-J. Elmers, G. Schönhense, M. Sing, and R. Claessen, Bulk Spin Polarization of Magnetite from Spin-Resolved Hard X-Ray Photoelectron Spectroscopy, *Phys. Rev.* B **104**, 045129 (2021).

[200] R. Winkler and M. Oestreich, *Physik Journal* **6** (2004).

[201] M. Ziese, *Rep. Prog. Phys*. **65**, 143 (2002).

[202] M. E. Fleet, *Acta Cryst*. B **37**, 917 (1981).

[203] H. Liu, G. Seifert, and C. Di Valentin, *J. Chem. Phys*. **150**, 094703 (2019).

[204] W. Wang, J.-M. Mariot, M. C. Richter, O. Heckmann, W. Ndiaye, P. De Padova, A. Taleb-Ibrahimi, P. Le Fèvre, F. Bertran, F. Bondino, E. Magnano, J. Krempaský, P. Blaha, C. Cacho, F. Parmigiani, and K. Hricovini, *Phys. Rev*. B **87**, 085118 (2013).

[205] S. F. Alvarado, M. Erbudak, and P. Munz, *Phys. Rev*. B **14**, 2740 (1976).

[206] S. F. Alvarado, W. Eib, F.Meier, D. T. Pierce, K. Sattler, H. C. Siegmann, and J. P. Remeika, *Phys. Rev. Lett*. **34**, 319 (1975).

[207] E. Kay, R. A. Sigsbee, G. L. Bona, M. Taborelli, and H. C. Siegmann, *Appl. Phys. Lett*. **47**, 533 (1985).

[208] H.-J. Kim, J.-H. Park, and E. Vescovo, *Phys. Rev*. B **61**, 15288 (2000).

[209] S. A. Morton, G. D. Waddill, S. Kim, I. K. Schuller, S. A. Chambers, and J. G. Tobin, *Surf. Sci*. **513**, L451 (2002).

[210] E. Vescovo, H.-J. Kim, J. M. Ablett, and S. A. Chambers, *J. Appl. Phys*. **98**, 084507 (2005).

[211] M. Fonin, R. Pentcheva, Y. S. Dedkov, M. Sperlich, D. V. Vyalikh, M. Scheffler, U. Rüdiger, and G. Güntherodt, *Phys. Rev.* B **72**, 104436 (2005).

[212] M. Fonin, Y. S. Dedkov, R. Pentcheva, U. Rüdiger, and G. Güuntherodt, *J. Phys.: Condens.* Matter **20**, 142201 (2008).

[213] D. J. Huang, C. F. Chang, J. Chen, L. H. Tjeng, A. D. Rata, W. P. Wu, S. C. Chung, H. J. Lin, T. Hibma, and C. T. Chen, *J. Magn. Magn. Mater*. **239**, 261 (2002).

[214] J. G. Tobin, S. A. Morton, S. W. Yu, G. D. Waddill, I. K. Schuller, and S. A. Chambers, *J. Phys.: Condens. Matter* **19**, 315218 (2007).

[215] Y. S. Dedkov, U. Rüdiger, and G. Güntherodt, *Phys. Rev*. B **65**, 064417 (2002).

[216] D. Schrupp, M. Sing, M. Tsunekawa, H. Fujiwara, S. Kasai, A. Sekiyama, S. Suga, T. Muro, V. A. M. Brabers, and R. Claessen, *Europhys. Lett*. **70**, 789 (2005).

[217] T. Pohlmann, T. Kuschel, J. Rodewald, J. Thien, K. Ruwisch, F. Bertram, E. Weschke, P. Shafer, J. Wollschläger, and K. Küpper, *Phys. Rev*. B **102**, 220411(R) (2020).

[218] M. Paul, M. Sing, R. Claessen, D. Schrupp, and V. A. M. Brabers, *Phys. Rev*. B **76**, 075412 (2007).

[219] L. Zhu, K. L. Yao, and Z. L. Liu, Phys. Rev. B **74**, 035409 (2006).

[220] R. Pentcheva, F. Wendler, H. L. Meyerheim, W. Moritz, N. Jedrecy, and M. Scheffler, *Phys. Rev. Lett*. **94**, 126101 (2005).

[221] M. Paul, D. Kufer, A. Müller, S. Brück, E. Goering, M. Kamp, J. Verbeeck, H. Tian, G. Van Tendeloo, N. J. C. Ingle, M. Sing, and R. Claessen, *Appl. Phys. Lett*. **98**, 012512 (2011).

[222] O. Fedchenko, A. Winkelmann, K. Medjanik, S. Babenkov, D. Vasilyev, S. Chernov, C. Schlueter, A. Gloskovskii, Y. Matveyev, W. Drube, B. Schönhense, H.-J. Elmers, and G. Schönhense, High-resolution hard-X-ray Photoelectron Diffraction in a Momentum Microscope - the Model Case of Graphite, *New J. Phys*. **21**, 113031 (2019).

[223] A. Winkelmann, B. Schröter, and W. Richter, *Phys. Rev*. B **69**, 245417 (2004).

[224] M. Paul, *Molecular Beam Epitaxy and Properties of Magnetite, Thin Films on Semiconducting Substrates*, Ph.D. Thesis, University of Würzburg (2010).

[225] Y. S. Dedkov, M. Fonin, D. V. Vyalikh, J. O. Hauch, S. L. Molodtsov, U. Rüdiger, and G. Güntherodt, *Phys. Rev.* B **70**, 073405 (2004).

[226] Z. Zhang and S. Satpathy, *Phys. Rev*. B **44**, 13319 (1991).

[227] C. Corder, P. Zhao, J. Bakalis, X. Li, M. D. Kershis, A. R. Muraca, M. G. White, and T. K. Allison, Ultrafast extreme ultraviolet photoemission without space charge, *Struct. Dyn*. **5**, 054301 (2018).

[228] C. Corder et al., Development of a tunable high repetition rate XUV source for time-resolved photo-emission studies of ultrafast dynamics at surfaces, *Proc.SPIE*, 10519 (2018).

[229] S. Chernov, J. Bakalis, A. Kunin, C. Corder, P. Zhao, M. G. White, G. Schönhense, and T. K. Allison, Excited state relaxation dynamics in HOPG using pump-probe momentum microscopy in perturbative limit, *AVS-67 book of abstracts,* publ. in preparation.

[230] A. Kunin, S. Chernov, J. Bakalis, C. Corder, P. Zhao, M. G. White, G. Schönhense and T. K. Allison, Electron Dynamics in $WS_2$ Probed by Time- and Angle-Resolved Photoemission Spectroscopy, *Faraday Discussion 'From optical to THz control of materials', book of abstracts,* publ. in preparation.

[231] D. Curcio, S. Pakdel, K. Volckaert, J. Miwa, S. Ulstrup, N. Lanata, M. Bianchi, D. Kutnyakhov, F. Pressacco, G. Brenner, S. Dziarzhytski, H. Redlin, S. Agustsson, K. Medjanik, D. Vasilyev, H.-J. Elmers, G. Schönhense, C. Tusche, Y.-J. Chen, F. Speck, T. Seyller, K. Bühlmann, R. Gort, F. Diekmann, K. Rossnagel, Y. Acremann, J. Demsar, W. Wurth, D Lizzit, L. Bignardi, P. Lacovig, S. Lizzit, C. E. Sanders, and P. Hofmann, Ultrafast electronic linewidth broadening in the C 1s core level of graphene, *Phys. Rev*. B **104**, L151104 (2021); doi: 10.1103/PhysRevB.104.L161104.

[232] K. Volckaert, *Ultrafast electronic and vibrational properties of Dirac materials*, Chapter 4: Ultrafast Lattice Vibrations in $Bi_2Se_3$, Ph.D. Thesis, Aarhus University, 2020.

[233] J. Madéo, M. K. L. Man, C. Sahoo, M. Campbell, V. Pareek, E. L. Wong, A. Al-Mahboob, N. S. Chan, A. Karmakar, B. M. K. Mariserla, X. Li, T. F. Heinz, T. Cao, and K. M. Dani, Directly visualizing the momentum-forbidden dark excitons and their dynamics in atomically thin semiconductors, *Science* **370**, 1199–1204 (2020).

[234] M. Schüler, U. De Giovannini, H. Hübener, A. Rubio, M. A. Sentef, T. P. Devereaux, and P. Werner, How Circular Dichroism in Time- and Angle-Resolved Photoemission Can Be Used to Spectroscopically Detect Transient Topological States in Graphene, *Phys. Rev*. *X* **10**, 041013 (2020).

[235] P. Kliuiev, G. Zamborlini, M. Jugovac, Y. Gurdal, K. von Arx, K. Waltar, S, Schnidrig, R, Alberto, M, Iannuzzi, V. Feyer, M. Hengsberger, J. Osterwalder, and L. Castiglioni, Combined orbital tomography study of multiconfigurational molecular adsorbate systems, *Nat. Commun*. **10**, 5255 (2019); doi: 10.1038/s41467-019-13254-7.

[236] A. Pancotti, J. Wang, P. Chen, L. Tortech, C.-M. Teodorescu, E. Frantzeskakis, and N. Barrett, X-ray photoelectron diffraction study of relaxation and rumpling of ferroelectric domains in BaTiO3(001), *Phys. Rev. B*, **87**, 184116 (2013).

[237] J. Maklar, S. Dong, S. Beaulieu, T. Pincelli, M. Dendzik, Y. W. Windsor, R. P. Xian, M. Wolf, R. Ernstorfer, and L. Rettig, A quantitative comparison of time-of-flight momentum microscopes and hemispherical analyzers for time and angle-resolved photoemission spectroscopy experiments, *Rev. Sci. Instrum*. **91**, 123112 (2020).

[238] H. Robinson, The secondary corpuscular rays produced by homogeneous X-rays, *Proc. Roy. Sec. Ser. A,* **104**, 455 (1923).

[239] O. Fedchenko, A. Winkelmann, and G. Schönhense, Structure Analysis using Time-of-Flight Momentum Microscopy With Hard X-rays: Status and Prospects, *submitted to J. Phys. Soc. Jap.*

[240] M. Puppin, Y. Deng, C. W. Nicholson, J. Feldl, N. B. M. Schröter, H. Vita, P. S. Kirchmann, C. Monney, L. Rettig, M. Wolf, and R. Ernstorfer, Time- and angle-resolved photoemission spectroscopy of solids in the extreme ultraviolet at 500 kHz repetition rate, *Rev. Sci. Instrum.* **90**, 023104 (2019).

[241] M. Keunecke, C. Möller, D. Schmitt, H. Nolte, G. S. M. Jansen, M. Reutzel, M. Gutberlet, G. Halasi, D. Steil, S. Steil, and S. Mathias, Time-resolved momentum microscopy with a 1 MHz high-harmonic extreme ultraviolet beamline, *Rev. Sci. Instrum*. **91**, 063905 (2020).

[242] S. V. Chernov, K. Medjanik, C. Tusche, D. Kutnyakhov, S. A. Nepijko, A. Oelsner, J. Braun, J. Minár, S. Borek, H. Ebert, H. J. Elmers, J. Kirschner, and G. Schönhense, Anomalous d-like surface resonances on Mo(110) analyzed by time-of-flight momentum microscopy, *Ultramicr*. **159**, 463 (2015).

[243] T. Duden and E. Bauer, A compact electron spin-polarization manipulator, *Rev. Sci. Instr*.**66**, 2861 (1995); doi: 10.1063/1.1145569.

[244] U. Heinzmann and J. H. Dil, Spin–orbit-induced photoelectron spin polarization in angle-resolved photoemission from both atomic and condensed matter targets, *Journal of Physics: Condensed Matter* **24**, 173001 (2012).

[245] K. Bühlmann, R. Gort, A. Fognini, S. Däster, S. Holenstein, N. Hartmann, Y. Zemp, G. Salvatella, T. U. Michlmayr, T. Bähler, D. Kutnyakhov, K. Medjanik, G. Schönhense, A. Vaterlaus, and Y. Acremann, Compact setup for spin-, time-, and angle-resolved photoemission spectroscopy, *Rev. Sci. Instrum*. **91**, 063001 (2020).

[246] R. Gort, K. Bühlmann, S. Däster, G. Salvatella, N. Hartmann, Y. Zemp, S. Holenstein, C. Stieger, A. Fognini, T. U. Michlmayr, T. Bähler, A. Vaterlaus, and Y. Acremann, Early Stages of Ultrafast Spin Dynamics in a 3d Ferromagnet, *Phys. Rev. Lett*. **121**, 087206 (2018).

[247] M. Keunecke, M. Reutzel, D. Schmitt, A. Osterkorn, T. A. Mishra, C. Möller, W. Bennecke, G. S. M. Jansen, D. Steil, S. R. Manmana, S. Steil, S. Kehrein, and S. Mathias, Electromagnetic dressing of the electron energy spectrum of Au(111) at high momenta*, Phys. Rev*. B **102**, 161403 R (2020).

[248] A. J. H. Boerboom, D. B. Stauffer, and F.W. McLafferty, Theory of the dodecapole, *Int. J. of Mass Spectr. and Ion Phys*. **63**, 17 (1985); doi 10.1016/0168-1176(85)87037-3.